\documentclass[prd,nofootinbib,twocolumn,preprintnumbers,superscriptaddress]{revtex4}

\pdfoutput=1

\usepackage{hyperref}

\usepackage{color}
\usepackage[dvipsnames]{xcolor} 
\usepackage[utf8x]{inputenc}
\usepackage[squaren]{SIunits}
\usepackage{amsmath}
\usepackage{amsfonts}
\usepackage{slashed}
\usepackage{graphicx}
\usepackage{subfigure}
\usepackage{rotating}
\usepackage{ulem}
\usepackage{enumitem}

\usepackage{graphicx,amsmath,amssymb,epsfig,verbatim,mathrsfs,array,layout,textcomp,latexsym}
\usepackage{multirow}

\allowdisplaybreaks[1]

\newcommand{\eq}[1]{\eqref{eq:#1}}
\newcommand{\fig}[1]{Fig.~\ref{fig:#1}}
\newcommand{\tab}[1]{Tab.~\ref{tab:#1}}

\newcommand{\refeq}[1]{Eq.~(\ref{eq:#1})}

\newcommand{\alphae}{\alpha_\mathrm{e}}

\newcommand{\bra}[1]{\left\langle{#1}\right\vert}
\newcommand{\ket}[1]{\left\vert{#1}\right\rangle}

\newcommand{\cl}[1]{\ensuremath{#1\,\%\,\text{C.L.}}}

\begin{document}
\preprint{DO-TH 22/25}

\setlength{\parindent}{0pt}

\title{
Two is better than one: The 
$\boldsymbol{U}$-spin-CP anomaly in charm}

\author{Rigo Bause}
\email{rigo.bause@tu-dortmund.de}
\author{Hector Gisbert}
\email{hector.gisbert@tu-dortmund.de}
\author{Gudrun Hiller}
\email{ghiller@physik.uni-dortmund.de}
\author{Tim~H\"ohne}
\email{tim.hoehne@tu-dortmund.de}
\affiliation{TU Dortmund University, Department of Physics, Otto-Hahn-Str.4, D-44221 Dortmund, Germany}
\author{Daniel~F.~Litim}
\email{d.litim@sussex.ac.uk}
\affiliation{Department of Physics and Astronomy, University of Sussex, Brighton, BN1 9QH, U.K.}
\author{Tom~Steudtner}
\email{tom2.steudtner@tu-dortmund.de}
\affiliation{TU Dortmund University, Department of Physics, Otto-Hahn-Str.4, D-44221 Dortmund, Germany}

\begin{abstract}
The recent measurement of the CP-asymmetry in the decay $D \to K^+ K^-$ by LHCb, combined with $\Delta A_{\text{CP}}$, evidences a sizable CP-asymmetry in $D \to \pi^+ \pi^-$ decays, which requires a dynamical enhancement of standard model higher-order contributions over tree-level ones by a factor of two. The data furthermore imply huge U-spin breaking, about 4-5 times larger than the nominal standard model one of $\lesssim 30 \%$ in charm. Enhanced breakdown of the two approximate symmetries points to models that violate U-spin and CP and disfavors flavor singlet contributions such as chromomagnetic dipole operators as explanations of the data. We analyze the reach of flavorful $Z^\prime$ models for charm CP-asymmetries. Models feature explicit U-spin and isospin breaking, allowing for correlations with $D \to \pi^0 \pi^0$ and $D^+ \to \pi^+ \pi^0$ decays with corresponding CP-asymmetries at a similar level and sign as $D \to \pi^+ \pi^-$, about $ {\cal{O}}(1-2) \cdot 10^{-3}$. Experimental and theoretical constraints narrow down the shape of viable models: anomaly-free models are leptophobic -- or at least electro- and muo-phobic -- with light $Z^\prime$ below ${\cal{O}}(20)$ GeV, and can be searched for in low mass dijets at the LHC, $\Upsilon$ and charmonium decays, and dark photon signatures. A $Z^\prime$ around $\sim 3$ GeV or $\sim (5-7)$ GeV can relieve the tensions in the $J/\psi \to \pi^+ \pi^-$ and $\psi^\prime \to \pi^+ \pi^-$ branching ratios with pion form factors from fits to Babar and JLab data, and simultaneously explain the charm CP asymmetries. Models also feature sizable branching ratios into light right-handed neutrinos or vector-like dark fermions, which can be searched for in $e^+ e^- \to$~hadrons + invisibles at Belle II and BESIII. Due to the low new physics scale dark fermions may induce an early Landau pole which requires UV-completion near the TeV-scale.
\end{abstract}
\maketitle

\section{Introduction}

The LHCb collaboration measured the CP-asymmetry in
$D^0 \to K^+ K^-$  decays~\cite{LHCb:2022vcc}
\begin{align} \label{eq:KKLHCB}
    A_{\text{CP}}(K^+K^-)&=(6.8\pm5.4\pm1.6)\cdot10^{-4}~,
\end{align}
where the first and second errors are statistical and systematic, respectively. Together with  the previous LHCb measurement \cite{LHCb:2019hro}
\begin{align} \label{eq:ACPLHCB}
\Delta A_{\text{CP}}\,&=A_{\text{CP}}(K^+K^-)-A_{\text{CP}}(\pi^+\pi^-)\,\\
& =\,(-15.4\pm2.9)\cdot 10^{-4}~, \nonumber
\end{align} 
LHCb performed a fit determining both direct CP-asymmetries~\cite{LHCb:2022vcc}
\begin{align}\label{eq:cp}
    a_{K^-K^+}^d\,&=\,(7.7\pm5.7) \cdot 10^{-4}~,\nonumber\\
    a_{\pi^-\pi^+}^d\,&=\,(23.2\pm6.1) \cdot  10^{-4}~,
\end{align}
with a correlation $\rho(a_{K^-K^+}^d,a_{\pi^-\pi^+}^d)=0.88$, and leading to $3.8\,\sigma$ evidence of CP-violation in $D^0\to\pi^+ \pi^-$ decays. This is puzzling for two reasons: First, the CP-asymmetry $a_{\pi^-\pi^+}^d$ is  larger than $|\Delta A_{\text{CP}}|$. 
Therefore, a standard model (SM) interpretation of the former needs even more dynamical enhancement of higher-order contributions $h$ over the tree-level amplitude $t$ to compensate the Cabibbo-Kobayashi-Maskawa (CKM) suppression  $  a_{\pi^-\pi^+}^{d \, \text{SM}} \sim 2 \cdot \text{Im}( V_{cb}^* V_{ub}/(V_{cd}^* V_{ud})) \, h/t \sim 1.2  \cdot 10^{-3}\, h/t$, with  data (\ref{eq:cp}) pointing to $h/t \sim 2$. 
Here an order one strong phase is assumed, and the enhancement is even bigger  if the latter  is suppressed, see App.~\ref{app:SM} for details.
Secondly, the new result implies a violation of $U$-spin symmetry, 
that is, violation of $a_{K^-K^+}^d =- a_{\pi^-\pi^+}^d$, at the level of $2.7\sigma$~\cite{LHCb:2022vcc}. 
Two approximate symmetries of the SM are thus being challenged. While it is too early to draw firm conclusions given the significant hadronic uncertainties in $D$-decays, 
the recent data make new physics (NP) searches with rare charm decays just more interesting -- this could be a hint for flavorful physics beyond the SM.

\begin{figure}[h!]
        \centering
      \includegraphics[scale=0.23]{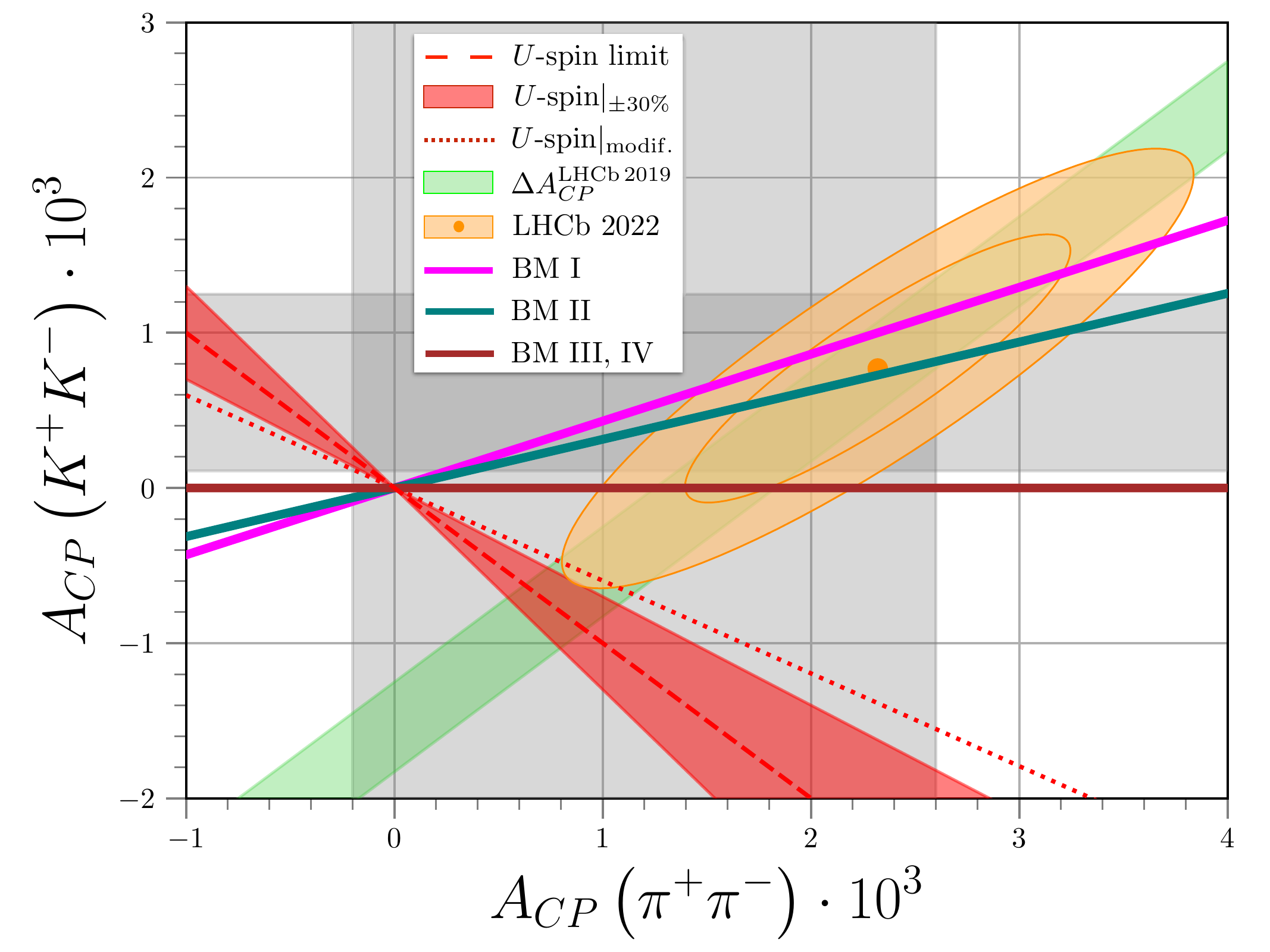}
        \caption{The $\boldsymbol{U}$-spin-CP anomaly in charm, showing bounds \eqref{eq:ACPLHCB} from LHCb on $\Delta A_{CP}$ (green-shaded area), bounds from \eqref{eq:cp} on  $A_\text{CP}(K^+K^-)$, and  $A_\text{CP}(\pi^+\pi^-)=(12\pm14)\cdot 10^{-4}$ from the Heavy Flavor Averaging Group (HFLAV)~\cite{HFLAV:2022pwe} (grey-shaded areas).   Values of $a_{K^-K^+}^d$ and $ a_{\pi^-\pi^+}^d$ from \eqref{eq:cp} are shown with correlation at 68 \% and 95 \% CL~\cite{LHCb:2022vcc}  (orange-shaded).  Also shown is the U-spin limit (red dashed line) together with $\lesssim30 \%$ SM-like breaking (red cones), and the modified U-spin relation \eqref{eq:modified} (red dotted line). Thick straight lines relate to the new benchmark models of this work: BM~I (magenta), BM~II (teal), BM~III (for $G=0$) and BM~IV (both brown),       see Table~\ref{tab:ZprimeModels}.
         }\label{fig:pi-K-plane}
\end{figure}
    
This interplay of the CP-asymmetries is illustrated in Fig.~\ref{fig:pi-K-plane}. 
The  small value of $A_{\text{CP}}(K^+K^-)$, combined with $\Delta A_{\text{CP}}$  (green-shaded area),  implies a sizable CP-asymmetry in $\pi^+  \pi^-$, together with substantial $U$-spin breaking, which has also been pointed out  in \cite{Schacht:2022kuj}.  
Predictions in the U-spin limit (red dashed line) and $\lesssim30 \%$ SM-like breaking (red-shaded cones) are indicated. 
The LHCb-fit (orange-shaded area) is two sigmas outside of this cone. The U-spin splitting in the $D \to  \pi^+  \pi^-$ and  $D \to  K^+  K^-$  branching ratios is very well-known~\cite{Savage:1991wu}, and  can be explained within the SM with $\lesssim30 \%$ breaking, for instance,   \cite{Pirtskhalava:2011va,Brod:2012ud,Hiller:2012xm}. Roughly speaking, because $\frac{(1+1/3)^2}{(1-1/3)^2}=4$ an assumed  33 \% contribution to both decays of opposite sign is more than enough to explain the enhancement of ${\cal{B}} (D \to  K^+  K^-)/{\cal{B}} (D \to  \pi^+  \pi^-)\simeq 2.8$, with or without considering the different phase space or factorizable flavor breaking from, {\it e.g.,} decay constants and form factors~\cite{Feldmann:2012js}. The splitting in the leading SM  decay amplitudes suggests a modified U-spin relation, see App.~\ref{app:SM},
\begin{equation}\label{eq:modified}
\frac{a_{K^-K^+}^d}{ a_{\pi^-\pi^+}^d}=-\sqrt{\frac{{\cal{B}} (D \to  \pi^+  \pi^-)}{{\cal{B}} (D \to  K^+  K^-)}}\,,
\end{equation}
also indicated in Fig.~\ref{fig:pi-K-plane} (dotted red line). Even though this effect slightly alleviates the anomaly, it still leaves the bulk of it unaltered, and the quest for models to explain it remains open.

Enhanced chromomagnetic dipole operators such as from supersymmetric loops are flavor singlets, feature therefore SM-like symmetries, and are not able to account for the significant U-spin breaking. Models that generically break flavor beyond the SM are $Z^\prime$ models with generation-dependent charges. Their impact on CP-asymmetries in charm has been studied in  \cite{Bause:2020obd}.

In this work, we analyze the new data  \eqref{eq:cp} within flavorful $U(1)^\prime$ extensions of the SM. 
Interestingly, due to  empirical constraints it  turns out that the  $Z^\prime$ mass has to be below the weak scale to induce a permille level CP-asymmetry in charm.
This is an important  finding and we plan to derive it step-by-step  in the paper:
$Z^\prime$ models and constraints from charm processes are discussed in Sec.~\ref{sec:Zp},  pointing to a low mass $Z^\prime$ of ${\cal{O}}(10) \, \text{GeV}$. In Sec.~\ref{sec:searches} we work out constraints applicable to this mass range from searches in dilepton and dijet signatures
and  quarkonium  decays.
In Sec.~\ref{sec:hi} we analyze the high energy behavior, including Landau poles. 
In Sec.~\ref{sec:con} we conclude.  
SM decay amplitudes  and observables are parametrized in App.~\ref{app:SM}. 
Details on the estimation of hadronic parameters are given in App.~\ref{app:A}. 
Charges for leptophobic, anomaly-free models with vanishing one-loop kinetic mixing are derived in App.~\ref{sec:newBM}, and their higher order kinetic mixing is studied in 
App.~\ref{sec:mix}.

\section{Flavorful \texorpdfstring{$Z^\prime$}{Z'} models \label{sec:Zp}}

We consider the SM extended by an abelian gauge group, with generation-dependent charges to fundamental fermions. We present the model set up in Sec.~\ref{sec:setup} and work out constraints from charm in Sec.~\ref{sec:bounds}. In Sec.~\ref{sec:CPpi0} predictions for CP-asymmetries in $D \to \pi^0 \pi^0$ and $D^+ \to \pi^+ \pi^0$ are given.

\subsection{\texorpdfstring{$Z^\prime$}{Z'} model set-up \label{sec:setup}}

We denote the $U(1)^\prime$-charges of the SM fermions $\psi=Q,U,D,L,E$ and and possibly also right-handed neutrinos $\nu_R$ as $F_{\psi_i}$, where $i=1,2,3$ corresponds to the generation label. The charges are subject to anomaly cancellation conditions \eq{ACC1}-\eq{ACC6}. The SM Higgs is uncharged under the $U(1)^\prime$ to avoid mixing with the electroweak sector. 
The theory has a rescaling invariance with a constant $k$ as $F_\psi \to k \,F_\psi, \, g_4 \to g_4/k$, where $g_4$ denotes the  $U(1)^\prime$-gauge coupling.  
It is therefore useful to consider rescaling invariant quantities such as $F_\psi \, g_4, \, F_\psi/F_{\psi^\prime}$ or $\text{d} g_4/g_4$. 
Here we choose to show integer charges for notational convenience.

\begin{figure}
    \centering
    \includegraphics[width=0.3\textwidth]{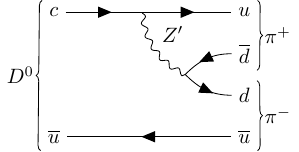} \\ [0.5cm]
    \includegraphics[width=0.3\textwidth]{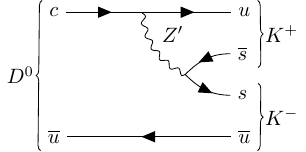}
    \caption{Contributions of the $Z^\prime$-boson to $D^0\to\pi^+\pi^-$ (top diagram) and $D^0\to K^+ K^-$ (bottom diagram) amplitudes.}
    \label{fig:tree}
\end{figure}

The $Z^\prime$ induces $D^0 \to \pi^+ \pi^-$ and  $D^0 \to K^+K^-$ at tree level, as illustrated in  Fig.~\ref{fig:tree}. Contributions to  the CP-asymmetries in \eqref{eq:cp} can be parametrized as~\cite{Bause:2020obd}~\footnote{The $Z^\prime$ induces also annihilation-type contributions  to the $D^0\to\pi^+\pi^-$ and $D^0\to K^+ K^-$ amplitudes. Annihilation contributions are power-suppressed and require gluon exchange; however, the actual size of suppressions in $D$-decays is within wider ranges  \cite{Grossman:2006jg}. We, therefore, refrain from including them in the numerical analysis, as we are focusing on the reach of models addressing data \eqref{eq:cp}. In addition, note that contributions induced by $F_{u_1}$ do not break U-spin.}
\begin{align}
    a_{K^-K^+}^d\,&=\, \frac{g_4^2}{M^2_{Z^\prime}} \Delta \widetilde{F}_R \left[c_K F_{Q_2} + d_K F_{d_2}\right],\nonumber\\
    a_{\pi^-\pi^+}^d\,&= \frac{g_4^2}{M^2_{Z^\prime}} \Delta \widetilde{F}_R \left[c_\pi F_{Q_1} + d_\pi F_{d_1}\right], \label{eq:cptheo}
\end{align}
where $c_{\pi,K},\,d_{\pi,K}$ are hadronic parameters (see App.~\ref{app:A}) and
\begin{align} \label{eq:delFR}
\Delta \widetilde{F}_R = \sin \theta_u\cos \theta_u\,( F_{u_2}-F_{u_1}) 
\end{align}
stands for the right-handed  $c \to u$ FCNC  coupling. As apparent, it requires non-universal charges $F_{u_2} \neq F_{u_1}$, as well as mixing between first and second generation right-handed up-quarks, described by the angle  $\theta_u$. 
We treat $\theta_u$ as a free parameter and adjust it accordingly. 
We assume that the corresponding angle in the down sector is sufficiently small to avoid kaon constraints. 
For the same reason, we consider only models with $F_{Q_{1,2}}=0$.~\footnote{This is also the reason why we do not consider scalar singlet mediators  contributing predominantly to $D^0\to\pi^+\pi^-$ decays: They would couple to left-handed (and right-handed) down quarks, and after CKM-mixing induce $\bar d s$-FCNCs, which are severely constrained.}
Explaining a sizable $a_{\pi^-\pi^+}^d$ and near or even SM-like $a_{K^- K^+}^d$ poses a challenge to  BSM model building. To estimate the maximal reach we assumed in \eqref{eq:cptheo}, which arises from interference between the SM and the $Z^\prime$ amplitudes, that the relative strong and  CP-phases are maximal. Having the latter near $\pi/2$ also evades constraints from CP-violation in $D$-mixing, see~\cite{Bause:2020obd} for details.

The efficiency of the BSM model in explaining  data  \eqref{eq:cp} is determined by the hadronic parameters $d_{K,\pi}, c_{K,\pi}$~\cite{Bause:2020obd}, of which here only $d_\pi,d_K$  matter. 
They include the leading order renormalization group (RG) running of Wilson coefficients in the weak effective theory from $M_{Z^\prime}$ to the charm mass scale, as well as the hadronic matrix elements. 
The latter are subject to sizable hadronic uncertainties~\cite{Grossman:2006jg,Brod:2011re,Altmannshofer:2012ur,Bause:2020obd}, see  App.~\ref{app:A} for details. 
The resulting CP-asymmetries serve rather as an indication of what is achievable in $Z^\prime$ models.

To construct models, that is, identify suitable charge assignments,  which account for the new LHCb results on CP-violation in charm  \eqref{eq:cp}, we follow similar lines as \cite{Bause:2020obd}:
Our starting point is  the cancellation of gauge anomalies,   decoupling from kaons $F_{Q_{1,2}}=0$, inducing an $c \to u$ FCNC $F_{u_2} \neq F_{u_1}$ and 
explicit $U$-spin breaking $F_{d_2} \neq F_{d_1}$. Absence of one-loop induced $Z$-$Z^\prime$-mixing is preferred.
In the following  the models are further narrowed down.
We discuss the theoretical and experimental constraints that arise and the corresponding selection criteria for charge patterns, which lead to the benchmark models \tab{ZprimeModels}.

 Let us also ask about the mass scale one would generically expect to address \eqref{eq:cp}  from  $Z^\prime$ -tree-level exchange.
Very roughly, assuming order one couplings, $g_4^2 \Delta \tilde F_R F_{d_1} \sim 1$, this gives a  $Z^\prime$ mass
around 
\begin{align} M_{Z^\prime} \sim (3  \sqrt{2} G_F V_{cd}^* V_{ud}   a_{\pi^-\pi^+}^d)^{-1/2}  \sim   7 \,  \mbox{TeV} \, , 
\label{eq:naive}
\end{align}
where RG-effects  in $d_\pi$ reduce this to the few TeV-range, see (\ref{eq:cptheo}),  App.~\ref{app:A} and \cite{Bause:2020obd}  for details.
In the next section,  we learn that the constraints from $D$-mixing require suppressed couplings, and a significantly lighter $Z^\prime$ than (\ref{eq:naive}).

\begin{table*}[t!]
	\centering
	\resizebox{0.8\textwidth}{!}{%
	\begin{tabular}{c|rcr|rcr|rcr|rcr|rcr|rcr}
	\hline \hline
		Model &&$F_{Q_i}$&&& $F_{u_i}$ &&& $F_{d_i}$ &&& $F_{L_i}$ &&& $F_{e_i}$ &&& $F_{\nu_i}$ &\\
		\hline
        BM I &0&	0&	0&	9&	-16&	7&	20&	-11&	-9&	15&	-6&	-9&	-16&	0&	16&	6&	12&	-18\\
        BM II & 0 & 0 & 0 & -19 & 9 & 10 & 20 & -8 & -12 & 4 & 1 & -5 & 15 & 2 & -17 & 8 & 2 & -10\\
        BM III & 0 &	0&	0&	$G$ &	 -$F$ &0	&	$F$&	-$G$&	0&	0&	0&	0& 0&	-$G$&	$F$&	0&	$G$&	-$F$ \\
        BM IV &0 &    0&    0&    -$F_u$ &   $ F_u$ &0   &    $F_d$&   0&  -$F_d$ &    0&    0&    0& $F_e$ &   0 & -$F_e$&       $F_\nu$&    -$F_\nu$&   0 \\
		\hline \hline
	\end{tabular}
	}	
	\caption{Benchmarks for  anomaly-free $U(1)^\prime$--extensions of the SM$+3\,\nu_R$. BM I, II  and IV avoid $Z-Z^\prime$ mixing at one loop, while BM III does not. Note, $|G/F| \ll1$ due to \eqref{eq:d2d1}, and $G=0$ is also possible. BM IV may or may not contain right-handed neutrinos, in which case $F_\nu=0$. It also can feature integer charges  with hierarchy $|F_e| \ll |F_u|, |F_d| $ in which case it becomes leptophobic, see App.~\ref{sec:newBM} for details and construction.
	Due to sizable couplings to electrons or muons the BMs I, II are excluded (\ref{eq:leptonlimit}) by $Z^\prime \to ee, \mu \mu$ searches for a light $Z^\prime$.
	While, in general, the ordering of generations is arbitrary due to permutation invariance, we use the ordering as stated here (the $i$th entry corresponds to the  $i$th 
	generation).  Note, BM III with only the charges in the right-handed up sector swapped, $F_{u_1}=-F, F_{u_2}=G$, is equally viable; we refer to it as BM III-s.}
		\label{tab:ZprimeModels}
\end{table*}

\subsection{Charming constraints \label{sec:bounds}}

We discuss constraints from charm CP-asymmetries, $D$-meson mixing, $D^0 \to \mu^+ \mu^-$ and on $\bar u c +\bar c u \to \ell^+ \ell^-$, $\ell=e,\mu,\tau$ from Dell-Yan production, as well as charm to invisibles.

\subsubsection{Charm CP-asymmetries}

Using Eqs.~\eqref{eq:cp} and \eqref{eq:cptheo}  with $F_{Q_{1,2}}=0$, the ratio between $F_{d_2}$ and $F_{d_1}$ is fixed,
\begin{align}\label{eq:d2d1}
    \frac{F_{d_2}}{F_{d_1}} = \frac{ d_\pi \,a_{K^-K^+}^d}{d_K\, a_{\pi^-\pi^+}^d} \simeq -\,0.42^{+0.83}_{-0.13}~,
\end{align}
resulting in a large hierarchy $|F_{d_2}| \ll  |F_{d_1}|$. 
The uncertainty in Eq.~\eqref{eq:d2d1} is computed from the $\chi^2$ function $\chi^2(a_{\pi^-\pi^+}^d,a_{K^-K^+}^d)$ with correlations included. 
This function can be expressed in terms of $a_{\pi^-\pi^+}^d$ (or $a_{K^-K^+}^d$) and the ratio $a_{K^-K^+}^d/a_{\pi^-\pi^+}^d$. We extract the uncertainty imposing $\Delta\chi^2=1$ and scanning $a_{\pi^-\pi^+}^d$ (or $a_{K^-K^+}^d$) within its 1$\sigma$ range. The non-parabolic behavior results in asymmetric uncertainties. Similar results were obtained in Ref.~\cite{Schacht:2022kuj}. 

Note that renormalization group equation (RGE) effects cancel in the ratio $d_\pi/d_K=-a_K/a_\pi \simeq -1.27\pm0.10$~\cite{Bause:2020obd}, therefore Eq.~\eqref{eq:d2d1} is independent of the $Z^\prime$ mass, and only a parametric dependence with the  quantities $a_{\pi,K}$ extracted from measured $D^0\to\pi^+\pi^-$ and $D^0\to K^+K^-$ branching ratios survive~\cite{Bause:2020obd}. Given the order of magnitude of  $a_{K^-K^+}^d$, within the ballpark of SM estimations, we also consider models with $F_{d_2}=0$.

\begin{figure*}[t!]
        \centering
        \includegraphics[width=0.5\textwidth]{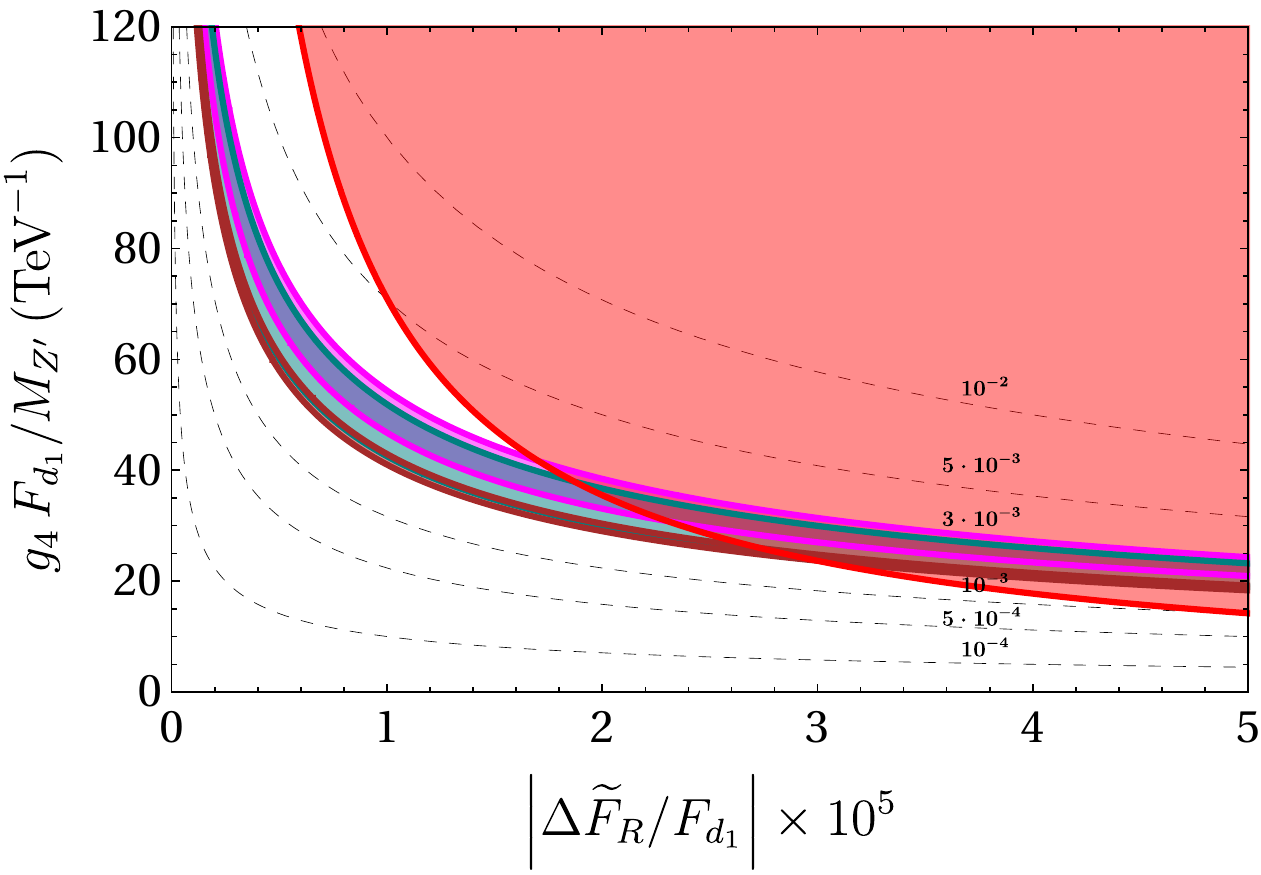}
        \caption{$g_4\,F_{d_1}/M_{Z^\prime}$ as a function of $\Delta\widetilde{F}_R/F_{d_1}$ and $d_\pi\simeq 0.1$TeV$^2$.  
        The red area represents the excluded \cl{95} region by $D$-mixing~\eqref{eq:mixing}. 
        The parameter space of models BM I, II, and III (with $G=0$) from Table~\ref{tab:ZprimeModels} accounting for the experimental results of $\Delta A_{\text{CP}}$ and $a^{d}_{\pi^-\pi^+}$ within its 1$\sigma$ range are shown in magenta, teal, and brown, respectively. BM IV has the same parameter space as BM III (with $G=0$). 
        The shaded bands include an additional 30\% of hadronic uncertainty. 
        The semi-analytical expressions of these regions are: $g_4 F_{d_1}/M_{Z^\prime}=c\,(\Delta\widetilde{F}_R/F_{d_1})^{-1/2}$ with factor $c=0.160\pm0.012\,\text{TeV}^{-1}$ (magenta), $0.149\pm0.015\,\text{TeV}^{-1}$ (teal), and $0.133\pm0.003\,\text{TeV}^{-1}$ (brown). The black dashed lines illustrate different values of $a^{d}_{\pi^-\pi^+}$.}
        \label{fig:constraints}
\end{figure*}

\subsubsection{\texorpdfstring{$D$}{D}-meson mixing}
   
$D$-meson mixing constrains right-handed up-quark couplings as
\begin{equation}\label{eq:mixing}
   \frac{g_4 \,\Delta \widetilde{F}_R}{M_{Z^\prime}} < 7.1 \cdot 10^{-4} \,\text{TeV}^{-1}\,\, (\cl{95}) \, , 
\end{equation}
where the right-hand side of this equation depends mildly  on the $Z^\prime$ mass  from RGE effects (it is a few percent for $M_{Z^\prime}\in[10,10^4]\,$GeV.) 
The limit~\eqref{eq:mixing} takes into account the recent update from HFLAV~\cite{HFLAV:mixing} where the new $D$-mixing experimental data from LHCb have been included~\cite{LHCb:2021ykz}. 
The bound for heavy $Z^\prime$ masses is somewhat stronger than the previous one, $8 \cdot 10^{-4}\,\text{TeV}^{-1}$~\cite{Bause:2020obd}. 

The available parameter space is presented in Fig.~\ref{fig:constraints} in a way that is independent of the $U(1)^\prime$ charge normalization.
Shown are curves in  $g_4\,F_{d_1}/M_{Z^\prime}$ versus  $\Delta\widetilde{F}_R/F_{d_1}$ that explain $a^{d}_{\pi^-\pi^+}$,
with uncertainties from data \eqref{eq:cp} which have been increased with an additional 30\% of uncertainty to account for hadronic effects.
Roughly,
\begin{align} \label{eq:c}
g_4\,F_{d_1}/M_{Z^\prime}  
\sqrt{\Delta\widetilde{F}_R/F_{d_1}} 
\sim 0.16 \, \text{TeV}^{-1}~.
\end{align}
In addition, the  excluded \cl{95} region by $D$-mixing~\eqref{eq:mixing} is shown in red. Thus,  $\Delta \widetilde{F}_R/F_{d_1}\ll 1$ via small mixing $\theta_u$ is instrumental to generate sizable CP-asymmetries while simultaneously avoiding $D$-mixing constraints. We recall that $\Delta\widetilde{F}_R$ \eqref{eq:delFR} contains the mixing angle $\theta_u$ which can be freely adjusted. We provide other (central) values of $a^{d}_{\pi^-\pi^+}$ as black dashed lines. We learn that the minimal value of $g_4\,F_{d_1}/M_{Z^\prime}$ with current data  is around $\sim 30\, \mathrm{TeV}^{-1}$, suggesting a low, sub-electroweak $Z^\prime$-mass.
\

\subsubsection{Charm dilepton and invisibles data}

Concerning charming dilepton  processes, the constraints from branching ratios of (semi-)muonic $D$--decays read~\cite{Gisbert:2020vjx,ICHEP22:LHCb}
\begin{align}
\begin{split}\label{eq:bounds_brsemimuonic}
 g_4^2 \vert \Delta \widetilde{F}_R\vert \sqrt{F^2_{L_2}+ F^2_{e_2}} \lesssim 0.02 \left(\frac{M_{Z^\prime} }{\rm TeV}\right)^2 ,
 \end{split}\\
 \begin{split}\label{eq:bounds_brmuonic}
 g_4^2 | \Delta \widetilde{F}_R (F_{L_2}- F_{e_2})| \lesssim 0.02 \left(\frac{M_{Z^\prime}}{\rm TeV}\right)^2 .
 \end{split}
\end{align}
Here, we  employed the recent LHCb measurement~\cite{ICHEP22:LHCb}, 
\begin{align}
    \mathcal{B}(D^0\to \mu^+\mu^-)<2.9\cdot 10^{-9} \,(\cl{90}),
\end{align}
which is a factor two stronger than the previous one. For $\ell=e,\tau$ Drell-Yan constraints \cite{Fuentes-Martin:2020lea} are stronger than those from rare decays 
\begin{align}
\begin{split}\label{eq:bounds_eand tau}
 g_4^2 \vert \Delta \widetilde{F}_R\vert \sqrt{F^2_{L_1}+ F^2_{e_1}} \lesssim 0.06  \left(\frac{M_{Z^\prime}}{\rm TeV}\right)^2 ,
 \end{split}\\
 \begin{split}
 g_4^2 | \Delta \widetilde{F}_R \vert \sqrt{F^2_{L_3}+ F^2_{e_3}} \lesssim 0.12  \left(\frac{M_{Z^\prime}}{\rm TeV}\right)^2  \, . 
 \end{split}
\end{align}
Using the relation \eqref{eq:c} imposed by $a^{d}_{\pi^-\pi^+}$, displayed in Fig.~\ref{fig:constraints}, the dilepton bounds are satisfied if
\begin{align}
\label{eq:rare-muon}
|F_{L_2}- F_{e_2}|, \sqrt{F^2_{L_2}+ F^2_{e_2}} &\lesssim 0.8\,|F_{d_1}| \,,  \\
\label{eq:rare-electron}
 \sqrt{F^2_{L_1}+ F^2_{e_1}} &\lesssim 2.3\,|F_{d_1}|\,,\\
 \sqrt{F^2_{L_3}+ F^2_{e_3}}&\lesssim 4.7\,|F_{d_1}| \, , 
 \label{eq:rare-tau}
\end{align}
that is, couplings to the leptons should not be excessive compared to the ones to the quarks. 

We also work out limits from data on  $c\to u$ plus missing energy.
Missing energy can stem from  right-handed neutrinos $\nu$ and/or vector-like dark BSM fermions $\chi$ charged under the $U(1)'$ only,
with mass not exceeding $m_D/2 \approx 0.9$ GeV.
We start with $D^0\to\pi^0+\text{invisibles}$, whose branching ratio is constrained by 
BES III data~\cite{BESIII:2021slf}
\begin{align} \label{eq:nunuBES}
    \mathcal{B}(D^0\to\pi^0\ \text{inv.})&< 2.1 \cdot 10^{-4} \,(\cl{90})\,.
\end{align}
Neglecting finite $m_\chi$ corrections,
the branching ratio can be written as~\cite{Bause:2020xzj,Bause:2020auq}
\begin{align}
    \mathcal{B}(D^0\to\pi^0\nu\bar{\nu},\chi\bar{\chi})&\approx\frac{2\pi^2 A_+}{G_F^2\alphae^2}\left(\frac{g_4^2 \Delta\widetilde{F}_R F_{\nu,\chi}}{M^2_{Z^\prime}}\right)^2\,,
\end{align}
where $A_+=9\cdot 10^{-9}$~\cite{Bause:2020xzj,Bause:2020auq}, and the sum over all flavors of the $\nu$ and the $\chi$ is understood. 
Following the previous analysis for the charged lepton constraints,
we obtain
\begin{align} \label{eq:nu1}
   |F_{\nu,\chi}| \lesssim 110 \, |F_{d_1}|\,.
\end{align}
Note the bound can be stronger if more than one kind or  flavor contributes.

The upper limit on $D^0\to\text{invisibles}$ by Belle~\cite{Belle:2016qek}
\begin{align}
    \mathcal{B}(D^0\to\text{inv.})&< 9.4 \cdot 10^{-5} \,(\cl{90})\,,
\end{align}
is in principal beneficial for massive invisibles (respecting  $m_{inv}< m_{D}/2$),
however does not constrain decays to fermions with purely vectorial coupling to the $Z^\prime$, such as  $g_4 F_\chi \bar \chi \gamma_\mu \chi$.

\subsubsection{Synopsis charm constraints and benchmarks}

Charm constraints imply further selection criteria on the model charges: $U$-spin breaking and hierarchy $F_{d1} \gg F_{d2}$ (\ref{eq:d2d1}),
on lepton couplings (\ref{eq:rare-muon})-(\ref{eq:rare-tau}), and on  invisible  and neutrino couplings (\ref{eq:nu1}).
All  benchmarks BM I -IV given in  Tab.~\ref{tab:ZprimeModels} pass these constraints.
Note that the BSM benchmarks  from \cite{Bause:2020obd} are disfavored by the new data. 
BMs I and II are obtained by scanning  integers. BMs III and IV are targeted towards more minimal models, with
BM IV designed to have no one-loop kinetic mixing.
BMs III and IV pass the additional constraints that arise from light $Z^\prime$ searches discussed in the next Sec.~\ref{sec:searches},
while BMs I and II fail to do so.
BM III-s, a variant of BM III  with the  charges between first and second generation up-type quark singlets swapped, $F_{u1}=-F$, $F_{u2}=G$ is equally  viable.
It has a different phenomenology than the other BMs, as it does not couple necessarily directly to charm quarks.

Indeed the main impact from $D$-mixing is that the mass of the $Z^\prime$ is  light, below the weak scale. 
Using  Eq.~\eqref{eq:cptheo} with $F_{Q_1}=0$, and the $D$-mixing bound 
we obtain a useful relation
\begin{align} \label{eq:g4mz}
   \frac{g_4 F_{d_1} }{M_{Z^\prime}} \sim    \frac{1}{ 0.025 \,  \text{TeV}} \times  \frac{|a_{\pi^-\pi^+}^d| }{ 0.002} \, , 
\end{align}
indicating a low NP mass scale,
 significantly lower than the naive estimate (\ref{eq:naive}) due to the severe constraints from Eq.~\eqref{eq:mixing}.
The ratio of coupling over mass required to explain $\Delta A_{\text{CP}}$  alone  \cite{Bause:2020obd} is approximately a factor of a few smaller 
 than the one from $a_{\pi^-\pi^+}^d$  \eqref{eq:cp} due to the smaller value of the CP-asymmetry, and cooperating contributions from both $KK$ and $\pi \pi$ asymmetries at least for modest U-spin breaking. The contribution of the flavorful $Z^\prime$ to four-quark operators $\bar u \,c \,\bar q\, q$, $q=d,s$ is about $2-3$ orders of magnitude smaller than
the one induced in the SM by $W$-exchange. Therefore, the $Z^\prime$-contribution is irrelevant  for the 
$D \to \pi^+ \pi^-$ and $D \to K^+  K^-$ branching ratios.

We observe that the anomaly-free models feature U-spin breaking and also isospin breaking, see Table~\ref{tab:ZprimeModels}. This implies signal in other 2-body charm CP-asymmetries, such as $\pi^+ \pi^0$, and  $\pi^0 \pi^0$, see \cite{Bause:2020obd}, and recently~\cite{Wang:2022nbm}. We work out predictions in Sec.~\ref{sec:CPpi0}.

\subsection{\texorpdfstring{$A_{\text{CP}}(\pi^0\pi^0)$}{ACP(pi0 pi0)} and \texorpdfstring{$A_{\text{CP}}(\pi^+\pi^0)$}{ACP(pi+pi0)} \label{sec:CPpi0}}

Flavorful  $Z^\prime$ models for $\Delta A_{\text{CP}}$ also induce CP-asymmetries in the $D \to \pi^0\pi^0$ and $D \to\pi^+\pi^0$ decays~\cite{Bause:2020obd}.
They are of similar size
\begin{align}
    \frac{A_{\text{CP}}(\pi^0\pi^0)}{A_{\text{CP}}(\pi^+\pi^0)}\,
    =\,\frac{d_{\pi^0}}{d_{\pi^\prime}}
    \simeq 1.08\pm0.10~,
\end{align}
and we recall that $A_{\text{CP}}(\pi^+\pi^0)$ requires  isospin violation to be finite.
Using  $F_{d_2} \ll F_{d_1}$,  
we find that  all CP-asymmetries involving pions are generically correlated as
\begin{align} \label{eq:correlation}
    A_{\text{CP}}(\pi^+\pi^0) \simeq  \frac{ d_{\pi^\prime}}{ d_{\pi^0}}   A_{\text{CP}}(\pi^0 \pi^0)  \simeq -\frac{ d_{\pi^\prime}}{ d_{\pi}}    \left( 1-\frac{F_{u_1}}{F_{d_1}}\right) \Delta A_{\text{CP}}
\end{align}
Since all $d_\pi$'s are roughly of the same size, and noting that viable benchmarks obey $|F_{u_1}| < |F_{d_1}|$ (see Tab.~\ref{tab:ZprimeModels} and Sec.~\ref{sec:searches}), 
the $Z^\prime$-induced CP-asymmetries are at the level of $\Delta A_{\text{CP}}$, which is a  permille. 
We also note the opposite sign of $\Delta A_{\text{CP}}$ with respect to  the others, hence $A_{\text{CP}}(\pi^0\pi^0)$ and $A_{\text{CP}}(\pi^+\pi^0)$ are positive in our models. 
Since $F_{u_1}/F_{d_1}$ can have either sign 
the relative factor $(1-F_{u_1}/F_{d_1})$ can be bigger or smaller than one. 
Concretely, it is $1$ for BM III (with $G=0$), 2 for the twisted BM III-s (with $G=0$) and within $1 \mp 1/\sqrt{2}$ for BM IV, depending on charges.
For the BM IV solution given by Eq.~\eqref{eq:BMVpart}, we obtain  a factor $1.7$.

\section{A flavorful \texorpdfstring{$Z^\prime$}{Z'} of the order 10 GeV? \label{sec:searches}}

Due to their strong impact on the viable mass range of the $Z^\prime$, we begin analyzing constraints from couplings to quarks in Sec.~\ref{sec:quark}.
The scale required to explain charm data~\eqref{eq:g4mz} points to a light $Z^\prime$. 
Searches for $U(1)^\prime$ extensions, including dark photons, $B-L$ and $B$-models in dileptons provide severe constraints in the 1-100 GeV range,
in particular in couplings to electrons and muons \cite{Ilten:2018crw}.
Consequently, couplings to electrons and muons, or leptons altogether should be 
 suppressed, much stronger  than in (\ref{eq:rare-muon})-(\ref{eq:rare-electron}). As such, BM I and II become excluded, and will  not be considered any further.
We quantify this and constraints in Sec.~\ref{sec:ZPrmumuee}, also working out couplings of the leptons that are induced by kinetic mixing. 
This effect is larger in BM III as kinetic mixing arises here already at 
one loop.
We analyse this and its impact in Sec.~\ref{sec:viable}.
In Sec.~\ref{sec:Zp-decay} we work out branching ratios of the $Z^\prime$.

\subsection{Mass constraints from $q \bar q$ \label{sec:quark}}

Constraints arise from dijets.
For $10\,\text{GeV} \lesssim M_{Z^\prime} \lesssim 50\,\text{GeV}$, 
the strongest constraints are from CMS~\cite{CMS:2019xai}, and their dijet plus initial state radiation (ISR) search~\cite{Dobrescu:2021vak}.  
Using their results, 
approximately $g_4 F_{d_1}\lesssim 0.5$, 
together with the constraint from charm~\eqref{eq:cp}, \eqref{eq:g4mz}
we arrive at the allowed mass range
\begin{align} \label{eq:searchrange}
10  \, \text{GeV} \lesssim M_{Z^\prime} \lesssim 20  \, \text{GeV}\,.
\end{align}

Around  and below $10\,\text{GeV}$,  constraints depend on the benchmark models.
Strong constraints from $\Upsilon \to  jj$ decays exist~\cite{Dobrescu:2014fca} around $10\,\text{GeV}$.
They apply to BM IV due to its $U(1)^\prime$ charge to $b$-quarks:
If one were to charge $s_R$ instead of $b_R$, the model would induce too large contributions to $D \to K^+ K^-$ decays.
Using  Ref.~\cite{Aranda:1998fr}, we obtain the allowed regions for BM IV with (\ref{eq:BMVpart})  from $\Upsilon(1s)$-decays respecting  CP-data~\eqref{eq:cp}, \eqref{eq:g4mz} as
\begin{align} \label{eq:upsilon}
M_{Z^\prime} \lesssim  7  \, \text{GeV}\,  \quad \text{or}  \quad
 M_{Z^\prime}  \gtrsim 15  \, \text{GeV}  \quad (\text{BM IV}) \, . 
\end{align}
On the other hand, BM III and BM III-s have no $Z^\prime$ coupling to $b$'s, and hence evade the $\Upsilon$-limits.

Charmonium decays  provide additional  constraints below $10\,\text{GeV}$ on BM III and BM IV, but not on the "swapped" model BM III-s, as it does not couple to charm (for $G=0$).
BM III-s with mass below (\ref{eq:searchrange}) can be probed in low energy hadronic processes involving first generation quarks, and invisibles.
Due to \eqref{eq:g4mz} the $Z^\prime$ below a GeV interacts feebly.
A detailed assesment of constraints and opportunities for  forward facilities \cite{Batell:2021snh} is below the scope of this work.

\begin{figure}
    \centering
    \includegraphics[width=0.3\textwidth]{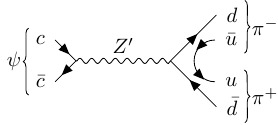} 
    \caption{Contribution of the $Z^\prime$  to $\psi \to\pi^+\pi^-$  amplitudes.}
    \label{fig:treepsi}
\end{figure}

We work out the  constraints on 
BM III and BM IV  from $\psi_i \to Z^{\prime*} \to \pi^+ \pi^-$  decays, $\psi_i=J/\psi, \psi^\prime$, with contributions illustrated in Fig.~\ref{fig:treepsi}. Following \cite{Bailey:1994qv}, we obtain for the branching ratios normalized to the ones into electrons,
\begin{align} \label{eq:psi}
&\frac{ {\cal{B}}(\psi_i \to \pi^+ \pi^-)}{{\cal{B}}(\psi_i \to e^+ e^-)} \frac{4}{|F_\pi(m_{\psi_i})|^2} =| 1+ A_{Z^\prime}/A_\gamma|^2 \, , \\
&\frac{A_{Z^\prime}}{A_\gamma} = \frac{ m_{\psi_i}^2}{ m_{\psi_i}^2 -M_{Z^\prime}^2 + i M_{Z^\prime} \Gamma(Z^\prime)} \frac{3 g_4^2 F_{u_2} (F_{u_1}-F_{d_1})}{8 \pi \alpha_e} \, ,
\end{align}
which depend on the ratio of  the $Z^\prime$-induced amplitude $A_{Z^\prime}$ to the SM-photon one $A_\gamma$, and the pion form factor $F_\pi$.
The  left-hand side  of  Eq.~(\ref{eq:psi})  is defined in such a way that by 
switching off the NP-amplitude it equals  one. 
We employ  the values of the pion form factor $|F_\pi(m_{J/\psi})|=0.056$,  $|F_\pi(m_{\psi^\prime})|=0.04$ from \cite{Cheng:2020vwr}, which uses  data on
$e^+ e^- \to \pi^+ \pi^-$ 
and pion-electron scattering as input.  As our models are electro-phob, we can safely assume that these data are not affected by the $Z^\prime$.
The $Z^\prime$-width $\Gamma(Z^\prime)$ is obtained from  (\ref{eq:gammaZp}).
\begin{figure}
        \centering
            \includegraphics[width=0.4\textwidth]{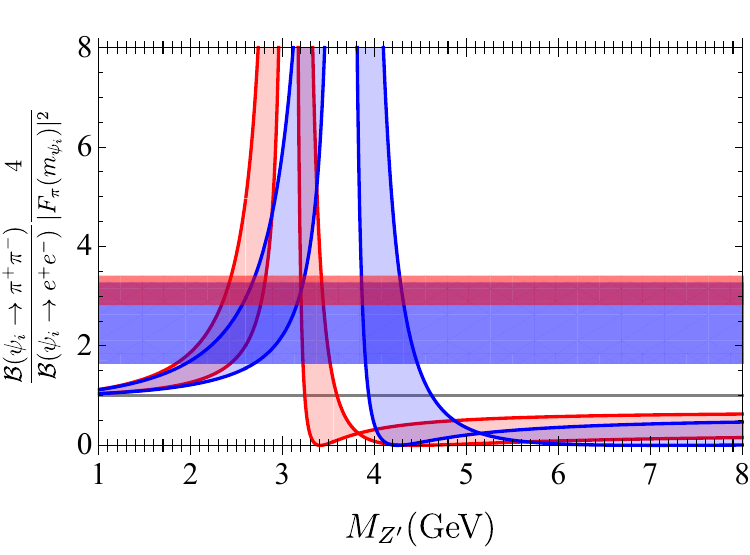}\\
          \includegraphics[width=0.4\textwidth]{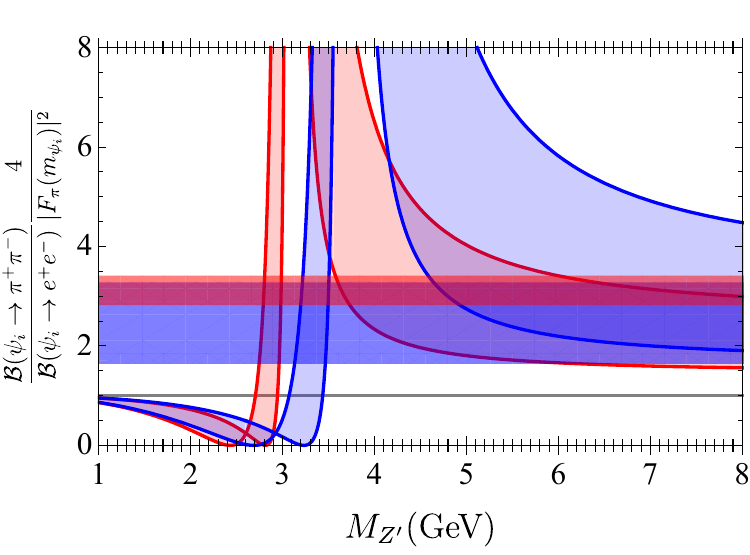}\\
        \caption{Constraints from charmonium decays.
        Horizontal red (blue)  bands denote the 
         left-hand side of (\ref{eq:psi})  from 1 sigma ranges of $J/\psi$-data with  $|F_\pi(m_{J/\psi})|=0.056$  ($\psi^\prime$ decays with 
        $|F_\pi(m_{\psi^\prime})|=0.04$.). Values of the pion form factor are from \cite{Cheng:2020vwr}.
        Curves correspond to the  predictions (right-hand side of (\ref{eq:psi}))  in BM III with $F \gg G$  (top) and BM IV with (\ref{eq:BMVpart}) (bottom)
              using \eqref{eq:g4mz}  including experimental uncertainties  from  (\ref{eq:cp}).
               The SM prediction via photon-exchange  is shown by the grey  line. }
                 \label{fig:psi}
\end{figure}

The constraints on BM III are shown in Fig.~\ref{fig:psi} (top), and  for BM IV with (\ref{eq:BMVpart}) (bottom).
The grey line denotes the SM prediction corresponding to  the ratios in (\ref{eq:psi}) being equal to one.
 One notices that the NP-contribution decouples very slowly for larger $Z^\prime$ masses, the reason is the growth of coupling with mass by means of  \eqref{eq:g4mz}.
We also include the  experimental uncertainties from the CP-data (\ref{eq:cp}).
The main difference between  the models BM III and BM IV  is stemming from  the $\gamma-Z^\prime$-interference term which has  opposite sign but similar size.
We observe  for BM III that  $Z^\prime$-masses around $[2.3,2.8]$ GeV  and  $[3.2,3.5]$ GeV are consistent with both the $J/\Psi$-data (red horizontal bands) and  the charm anomalies (red curves).
The ranges obtained using $\psi^\prime$-data (blue horizontal bands), which have larger uncertainties, are  $[1.9,3.3]$ GeV  or $[3.9,4.5]$ GeV. 
BM IV can explain charm CP-data and $J/\psi \to \pi \pi$ decays for  masses within $[2.8,3.0]$ GeV, or starting from $3.6$ GeV  until the 
$\Upsilon$-limit (\ref{eq:upsilon}) kicks in, at about $7$ GeV.
The corresponding ranges from the  $\psi^\prime$-data read $[3.1,3.5]$ GeV, or $[4.6,7]$ GeV.
Assuming compatibility with both charmonia \cite{ParticleDataGroup:2022pth},  $\Upsilon$ and charm CP data determines the  $Z^\prime$-mass, depending on the model, as
\begin{align}
\label{eq:det}
M_{Z^\prime} \sim \,  &[2.3,2.8]\,  \text{GeV} \,  ~\text{or}  ~[3.2,3.3] \, \text{GeV} \,  (\text{BM  III})\, ,  \\  \nonumber
M_{Z^\prime} \sim \,  &[4.6, 7 ]\, \text{GeV} \quad (\text{BM IV})  \, .  \nonumber
\end{align}
With these parameters our models provide an opportunity to resolve the longstanding  tension  between $F_\pi$ extracted from $J/\psi$-decays assuming the leading  photon-exchange contribution and  QCD, e.g.~\cite{Bruch:2004py}.
Note that the explanation of charmonia in BM III is a resonant effect, while in BM IV it is in  the tails.

We recall that our results are subject to sizable uncertainties:
the weak effective theory  is challenged since a  $Z^\prime$ as light as a few GeV  is  close to the charm scale, 
in addition to the uncertainties from hadronic matrix elements. We also neglect $G$-parity violating contributions  in the SM to the
charmonia decays, e.g. \cite{BaldiniFerroli:2016mbs}, noting that  the tension is significant, $7 \sigma$ for the $J/\pi$ and $1.8 \sigma$ for the $\psi^\prime$.
As such, we consider the phenomenology and formal constraints also in wider viable regions of $M_{Z^\prime}$.

For BM III, additional constraints from charmonia to taus or invisibles exist.
Using a similar computation as in (\ref{eq:psi}), we find that  ${\cal{B}}(\psi^\prime \to \tau^+ \tau^-)$ \cite{ParticleDataGroup:2022pth} 
gives the allowed  mass ranges $M_{Z^\prime}  \lesssim 2.2 $  GeV or   within $4.0 -4.8 $ GeV,  
very close to the windows implied by the pion form factor (\ref{eq:det}).  
In view of the large uncertainties further analyses are promising and desirable.
Furthermore,  ${\cal{B}}(\psi \to \text{nothing}) < 7 \cdot 10^{-4}$ \cite{ParticleDataGroup:2022pth}  
requires  either $M_{Z^\prime}  \lesssim 0.7 $  GeV, which is in conflict with (\ref{eq:det}), or the BSM neutrino which couples to the $Z^\prime$ to be
heavier than half the $\psi$ mass to forbid the decay kinematically.
This suggests that this benchmark solution to the charm CP-data can be probed in charmonium decays.

\subsection{\texorpdfstring{$Z^\prime \to e e$}{Z' -> ee} and \texorpdfstring{$\mu \mu$}{mu mu} bounds }
\label{sec:ZPrmumuee}

We work out constraints from  $Z^\prime\to e^+e^-,\mu^+\mu^-$ decay searches. First, we study models where the $Z^\prime$ couples directly to electrons or muons~\cite{Bause:2021prv,Fabbrichesi:2020wbt} such as BM I, II, III (with $G\neq 0$), and IV,
\begin{align}\label{eq:couplingepsilon}
    g_4\,F_{e_i,L_i}\,=\,\sqrt{8\pi\,\alphae}\,\varepsilon\,\approx 0.4\,\varepsilon \, . 
\end{align}
The experimental search limits are given in terms of the mixing parameter $\varepsilon$, defined as $\mathcal{L}_\varepsilon=-\varepsilon\,e\,J^\mu\,Z^\prime_\mu$ where $J^\mu$ is the electromagnetic current of SM fermions. For the range of interest~\eqref{eq:running-eps}, the current experimental limit on $\varepsilon$ both for electrons and muons is~\cite{Ilten:2018crw,LHCb:2019vmc}
\begin{align}\label{eq:leptonsepsilon}
    |\varepsilon(M_{Z^\prime})|\lesssim 10^{-3}~.
\end{align}
Combining Eq.~\eqref{eq:couplingepsilon} with \eqref{eq:leptonsepsilon}, one gets
\begin{align}\label{eq:g4Fl}
    g_4\,F_{e_{1,2},L_{1,2}}\lesssim 4\cdot 10^{-4}~.
\end{align}
Inspecting Fig.~\ref{fig:constraints}, we observe that $g_4\,F_{d_1}\gtrsim 0.3$ for $M_{Z^\prime}\gtrsim 10\,$GeV, which in combination with Eq.~\eqref{eq:g4Fl} leads to
a strong suppression of electron and muon couplings over the down-type quark one,
\begin{align}\label{eq:leptonlimit}
    \frac{F_{e_{1,2},L_{1,2}}}{F_{d_1}}\lesssim \frac{1}{750}
~,\end{align}
significantly stronger then the rare $D$-decay and Drell-Yan constraints \eqref{eq:rare-muon} and \eqref{eq:rare-electron}. 
Note that Eq.~\eqref{eq:leptonlimit} directly excludes BM I and II, and dictates a strong quark and lepton charge hierarchy in BM III (as $ |G| \lesssim 1.3 \cdot 10^{-3} \, |F|$) and IV. 

Next we study effects from kinetic mixing.
If the $Z^\prime$ does not couple directly to electrons and muons as in  BM III with $G=0$, one can still induce a small coupling $\varepsilon$ to $\mathcal{L}_\varepsilon$ from $Z^\prime-\gamma$ gauge-kinetic mixing 
\begin{equation}\label{eq:gauge-kinetic-mixing}
    \mathcal{L} \supset - \tfrac{1}{4} F^{\mu\nu}F_{\mu\nu}  - \tfrac{1}{4} Z^{\prime \mu\nu} Z^\prime_{\mu\nu} - \tfrac{\eta}{2} F^{\mu\nu} Z^\prime_{\mu\nu}\,,
\end{equation}
which yields 
\begin{equation}\label{eq:eps-eta}
    \varepsilon = - \frac{\eta}{\sqrt{1 - \eta^2}}\,.
\end{equation}
Note that gauge-kinetic mixing is in general not technically natural and cannot be switched off at more than one scale. It is also related to the gauge-kinetic mixing between the $Z'$ and the hypercharge field $B$ before electroweak symmetry breaking via $ \eta = \eta_{B-Z'} \cos \theta_W $. This implies $Z-Z'$ mass mixing generating a correction $\delta M_Z$ to the unmixed tree-level mass of the $Z$ boson, which affects the  $\rho$ parameter $\rho^{-1} = (M_Z + \delta M_Z )\cos \theta_W /M_W $: 
\begin{equation}
    \left(\frac{ \rho-\rho_{\text{SM}}}{\rho}\right) =  -\frac{\varepsilon^2 \tan^2 \theta_W }{2(1+ \varepsilon^2)} \left(\frac{M_{Z'}}{M_Z}\right)^2  + \mathcal{O}\left(\frac{M_{Z'}}{M_Z}\right)^4 \, ,  \\
\end{equation}
which for light $Z^\prime$'s is negative, but vanishes quadratically with $\varepsilon$. Here, $\rho_{\text{SM}}$ denotes the  $\rho$ parameters SM value, which is close to one.
The global fit of electroweak precision parameters  \cite{ParticleDataGroup:2022pth} suggests a relative NP contribution as
 \begin{equation}\label{eq:rho-NP}
    \left(\frac{\rho-\rho_{\text{SM}}}{\rho}\right) = (3.8 \pm 2.0) \cdot 10^{-4}\,.
\end{equation}
Thus, the light $Z'$ contributes with the opposite sign. However, the correction is within $2\sigma$ of \eq{rho-NP} if 
\begin{equation}\label{eq:eta_bound}
\left|\varepsilon(M_Z)\right| \lesssim 4 \cdot 10^{-1}  ~~~(7 \cdot 10^{-2} )
\end{equation}
for a $Z'$ mass above $3$~GeV ($15$~GeV). This has to be compatible with the constraint \eqref{eq:leptonsepsilon} at the $M_{Z'}$ scale.  As the running of $\varepsilon$ is in general not technically natural, avoiding both constraints may results in highly non-trivial conditions of all $U(1)'$ charges for a light $Z'$. In the following section, we discuss benchmark models that are feasible in this regard.

\subsection{Viable scenarios \label{sec:viable}}

We consider the benchmarks BM III and BM IV, see Tab.~\ref{tab:ZprimeModels}, which allow for sizable couplings to quarks but not leptons \eqref{eq:leptonlimit}. Note that in these models the top quark has no direct coupling to the $U(1)^\prime$.
This is also beneficial in suppressing  contributions from kinetic mixing.

\begin{itemize}
\item[A)]  BM IV follows the constructions of App.~\ref{sec:newBM}, with $|F_e| \ll  |F_{u,d}|$. The kinetic mixing is natural in this model and can be switched off or made feebly small,
see App.~\ref{sec:mix}.
\item[B)] BM III is electro- and muo-phobic. It has couplings to taus and $\nu_R$. The RG evolution of the kinetic mixing parameter reads
\begin{align}\label{eq:running-eps}
    \varepsilon(\mu) &= \varepsilon(\mu_0) - \delta_\varepsilon \,\ln\left(\frac{\mu}{\mu_0}\right) + \text{2-loop}\,,\\
    \delta_\varepsilon &= \frac{e g_4 F }{3 \pi^2} + \mathcal{O}(\varepsilon)\,.
\end{align}
As 
\begin{equation}\label{eq:BMIII-constraint-g4}
    g_4(M_{Z'}) F \gtrsim \frac{30\, M_{Z'}}{\text{TeV}},
\end{equation}
see Fig.~\ref{fig:constraints}, we roughly find a running 
\begin{equation}\label{eq:running-eps-Z-Z'}
 \begin{aligned}
    |\varepsilon(M_Z) - \varepsilon(M_{Z'})| &\gtrsim\frac{10\,e}{\pi^2} \,\frac{M_{Z'}}{\text{TeV}} \ln\left( \frac{M_{Z}}{M_{Z'}}\right) \\
    & \gtrsim 10^{-3}
 \end{aligned}
\end{equation}
between the $Z'$ and the electroweak scale, where in the last line we used $M_{Z^\prime} \gtrsim  3 \, \text{GeV}$.
Thus, the running can accommodate both $\rho$ parameter and $Z'\to \ell \ell$ constraints \eqref{eq:leptonsepsilon} and \eqref{eq:eta_bound} if $|\varepsilon(M_{Z})|\sim \mathcal{O}(10^{-2})$. 
Using the four-loop running~\cite{Bednyakov:2021qxa,Davies:2021mnc}, we  verified that the approximation \eqref{eq:running-eps} holds well for the lower end of \eqref{eq:BMIII-constraint-g4}. Larger values of $g_4(M_{Z'})F$ may result in \eqref{eq:running-eps-Z-Z'} increasing in the order of magnitude, eventually spoiling compatibility with the kinetic mixing constraints.

\end{itemize}

To summarize, bounds from $ee$ and $\mu \mu$ can always be evaded: In BM IV, the hierarchy between quark and lepton charges can simply be chosen larger.
In BM III, while kinetic mixing induces couplings to SM fermions that are uncharged before going to the gauge boson mass basis, the impact of this  can be avoided
by tuning with the contribution at the matching scale \eqref{eq:running-eps}, at the level of $0.1$. If experimental dilepton  constraints on $\varepsilon$ improve in the future, this can be accommodated with an increase of tuning
at similar level.  Of course, in a UV model this choice is not possible, therefore, models of type BM III can be just  around the corner and
show up in the next round of dilepton searches.
 
\subsection{\texorpdfstring{$Z^\prime$}{Z'} decay \label{sec:Zp-decay}}

\begin{table*}
	\centering
	\resizebox{0.8
	\textwidth}{!}{
	\begin{tabular}{l|c|c|c|c|c|c|c}
	\toprule 
	Model & light quarks & $\quad b\quad$ & $\quad c \quad$ & $\quad e \quad$ & $\quad \mu \quad$ & $\quad \tau \quad$ & $\quad \nu_R \quad$ \\ \hline
BM III$|_{M_{Z'}=2.5\, {\rm GeV}}$ \ \ \quad\quad\quad\quad\quad& $75$ & 0 & 0 & 0 & 0 & 0 & $25$\\
BM III$|_{M_{Z'}=15\, {\rm GeV}}$ \ \ \quad\quad\quad\quad\quad& $38$ & 0 & $37$ & 0 & 0 & $12$ & $13$\\  
BM III-s$|_{M_{Z'}=2.5\, {\rm GeV}}$ \ \ \quad\quad\quad\quad\quad& $86$ & 0 & 0 & 0 & 0 & 0 & $14$\\
BM III-s$|_{M_{Z'}=15\, {\rm GeV}}$ \ \ \quad\quad\quad\quad\quad& $75$ & 0 & 0 & 0 & 0 & $12$ & $13$\\   
BM IV$|_{M_{Z'}=5\, {\rm GeV}}$& $\quad 79 \quad$ & 0 & $21$ & 0 & 0 & 0 & 0 \\
BM IV$|_{M_{Z'}=15\, {\rm GeV}}$& $\quad 54 \quad$ & $28$ & $18$ & 0 & 0 & 0 & 0 \\
		\hline \hline
	\end{tabular}}
\caption{Tree-level branching fractions in \% for the different $Z^\prime$ decay modes to fermion-antifermion pairs. Results for BM III  and BM III-s are given in the limit $G \ll F$. In BM IV, branching ratios depend on the different charge assignments $F_{u,d,e,\nu}$, see main text for details. The branching ratios shown in this table are obtained from $F_u=985,\, F_d=1393,\,F_e=1$ in \eq{BMVpart} and $F_\nu=0$. Branching ratios in all BMs differ perceptibly between the low and high $M_{Z^\prime}$ windows, \eq{searchrange} and \eq{det}, as the decays $Z^\prime\to b\bar{b},\,c\bar{c},\,\tau^+\tau^-$ are kinematically forbidden or suppressed in the few GeV range. Corrections to branching ratios from kinetic mixing are generically $\lesssim 10^{-7}$.}
\label{tab:Brs}
\end{table*}

In this section, we work out branching ratios of the light $Z^\prime$ boson. 
The partial decay width of the $Z'$ to fermions $\psi$ with mass $m_{\psi} < M_{Z'}/2$ is given as \cite{Bause:2021prv}
\begin{align} \label{eq:gammaZp}
    & \Gamma(Z' \to \psi\bar{\psi}) = \frac{N_C^{\psi} g_4^2}{24\pi} M_{Z'}\sqrt{1-4\frac{m^2_{\psi}}{M^2_{Z'}}} \\
    & \cdot \left[ F^2_{\psi_L} + F^2_{\psi_R} -\frac{m^2_{\psi}}{M^2_{Z'}} (F^2_{\psi_L} -6 F_{\psi_L} F_{\psi_R} + F^2_{\psi_R}) \right],\nonumber
\end{align}
with color factor $N_C^{\psi}=3$ for quarks and $N_C^{\psi}=1$ otherwise. $F_{\psi_{L(R)}}$ denotes the $U(1)'$ charge of the left-handed (right-handed) fermion $\psi$. 
Due to the low mass of the $Z^\prime$ \eqref{eq:searchrange} decay channels  into the SM electroweak gauge bosons or the Higgs  are kinematically forbidden. 
For low $M_{Z^\prime} \lesssim \,\text{few}\times\text{GeV}$, also the fermionic decays $Z^\prime \to b\bar{b},\,c\bar{c},\,\tau^+ \tau^-$ can be either kinematically forbidden or severely phase space suppressed.
In the limit $m^2_{\psi}/ M^2_{Z'} \ll1$ on the other hand the branching ratios are simply given by
\begin{equation} \label{eq:ZPrwidth}
    \mathcal{B}(Z' \to \psi \bar{\psi}) = \frac{N_C^{\psi}(F^2_{\psi_{L}} + F^2_{\psi_{R}})}{
    \sum_{\psi'}N_C^{\psi'}(F^2_{\psi'_{L}} + F^2_{\psi'_{R}})
    }.
\end{equation}
Numerical results for branching ratios in BM III, III-s and IV are shown in \tab{Brs} for different $M_{Z^\prime}$. In BM III and III-s results are given in the limit $G \ll F$ where dimuon bounds are avoided.
For $M_{Z^\prime}=20, 10$ and $3$ GeV using (\ref{eq:g4mz}) in BM III we obtain the width $\Gamma(Z')=\sum_i \Gamma(Z' \to \psi_i \bar{\psi}_i) \theta (M_{Z^\prime} - 2 m_{\psi i})=1.8, 0.2$ and $4 \cdot 10^{-3}$ GeV, respectively. Very similar values for $\Gamma(Z^\prime)$ are found in BM III-s and IV. We note that for the lower masses Eq.~(\ref{eq:gammaZp}) is not accurate as hadronic final states should rather be  taken into account. 
In BM IV, results depend on the charge assignments $F_{u,d,e,\nu}$. However, dilepton bounds suggest suppression of lepton couplings \eqref{eq:leptonlimit}. When setting $|F_e|,|F_\nu| \ll |F_{u,d}|$ 
we asymptotically approach a leptophobic model, with decays only to $b,\,c$ and jets, {\it i.e.,} light quarks.
We also provide branching ratios for the concrete scenario \eqref{eq:BMVpart} in \tab{Brs}.
All benchmark models lead to a promptly decaying $Z^\prime$.

\begin{figure}[h!]
    \centering
    \includegraphics[width=0.4\textwidth]{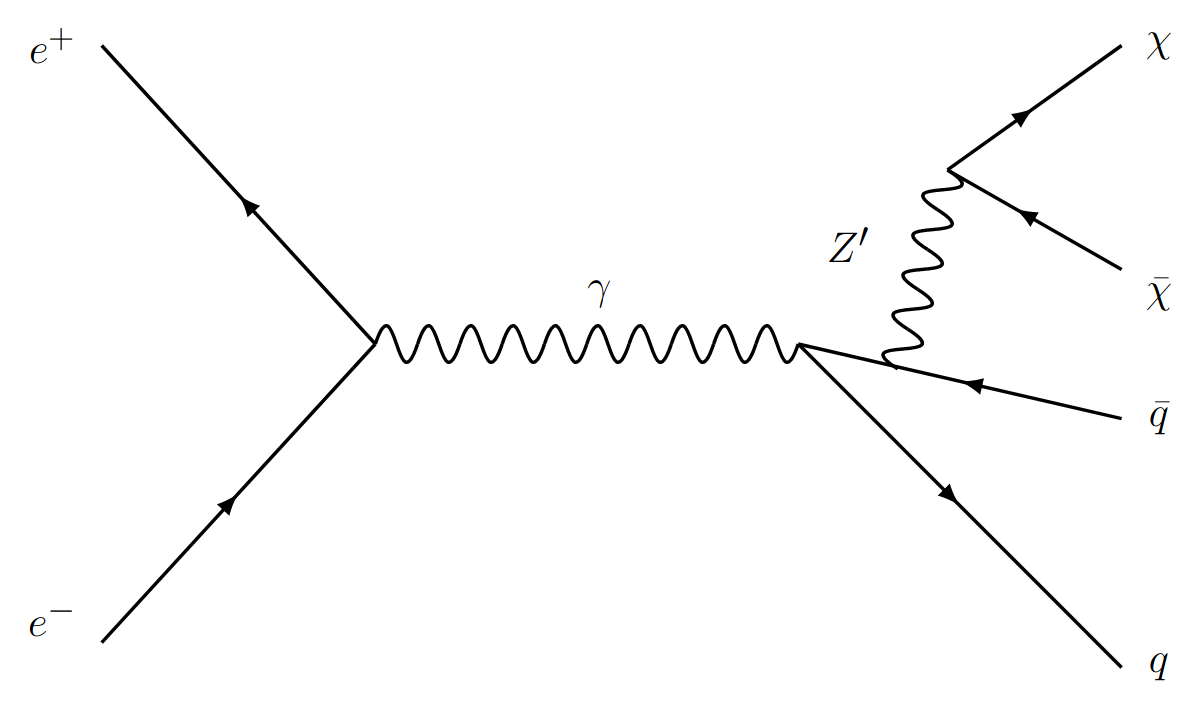}
    \caption{Smoking gun signature of  benchmark models at $e^+e^-$-machines for  $|F_{\nu/\chi}| \gtrsim |F_\psi|$ where $\psi=Q,\,u,\,d,\,L,\,e,\,\nu$. A $q\bar{q}$-pair is produced via an $s$-channel photon and radiates of a $Z'$ invisibly decaying to $\chi \bar{\chi}$ or $\nu \bar{\nu}$, 
    leading to a final state containing hadrons in association with missing energy. }
    \label{fig:missing}
\end{figure}

Another possibility to suppress branching ratios to quarks and charged leptons is  $|F_\nu| \gg |F_{u,d,e}|$. 
In this case the $Z'$ boson decays mostly invisibly to right-handed neutrinos. 
The same effect can also be achieved in all BMs by adding a light and dark vector-like BSM fermion $\chi$ with $U(1)'$ charge $F_\chi$. Due to its vector-like nature $\chi$ does not contribute to any gauge anomalies and it can have a simple Dirac mass term. Assuming $m_\chi < 2 M_{Z^\prime}$ as well as $|F_{\chi}| \gtrsim |F_\psi|$ for $\psi=Q,\,u,\,d,\,L,\,e,\,\nu$ the $Z^\prime$ will decay predominantly invisibly as $Z^\prime \to \chi \bar{\chi}$, see \eq{ZPrwidth}. 
For a heavy $Z^\prime$ this possibility has already been explored in the context of the $B$-anomalies \cite{Bause:2021prv}.
The $Z^\prime$ can be radiated off quarks, which in the above scenario leads to characteristic signatures such as hadrons in association with invisibles, i.e. missing energy, see \fig{missing}. 
Unfortunately, to our knowledge in the mass range of our interest there is no experimental analysis for this process available. 
However, an invisibly decaying $Z^\prime$ radiated off final state hadrons would be the smoking gun signature of this scenario at $e^+e^-$ machines potentially giving rise to tight bounds on $F_{\nu,\chi}$. 
This is in contrast to bounds from existing searches for $e^+e^- \to \gamma_\text{ISR} + (Z' \to \chi \bar{\chi})$ \cite{Campajola:2021pdl,BESIII:2022oww,BaBar:2017tiz} where cross sections are suppressed by tiny factors $\varepsilon^2$ or $F_e^2$ even if $\mathcal{B}(Z' \to \chi \bar{\chi}) \simeq 100\%$. 
However, note that $F_{\nu,\chi}$ cannot be arbitrarily large, as by adding additional $U(1)'$-charged matter we always increase the RG growth of $g_4$ which finally might give rise to a low energy Landau pole, see Sec. \ref{sec:hi} for details.
There are also constraints \eqref{eq:nu1} on the charges from rare decays for fermions lighter than $m_D/2$.

We comment on corrections to the $Z^\prime$ decay rate to SM fermions
from kinetic mixing. As discussed in \ref{sec:ZPrmumuee} the $Z'$ thus couples to the electromagnetic current $J^\mu$ via $\mathcal{L}_\varepsilon = - \varepsilon e J^\mu Z'_\mu$, where the kinetic mixing parameter $\varepsilon$ is defined in \eq{gauge-kinetic-mixing}, \eq{eps-eta}. The corresponding partial decay width can be obtained from \eq{ZPrwidth} by replacing $g_4 \to \varepsilon \,e$ and the $U(1)'$ charges with electric ones, $F_{\psi_{L,R}} \to q_{\psi}$. Experimentally, kinetic mixing is constrained by \eq{leptonsepsilon} to be small. Thus, decays via kinetic mixing are suppressed in comparison to unmixed decays by a tiny factor 
\begin{equation}
   \kappa \sim \frac{\sum_{\psi} (2 \, N_C^\psi \, q_\psi^2) \varepsilon^2 \, e^2}{\sum_{\psi} N_C^{\psi}(F^2_{\psi_{L}} + F^2_{\psi_{R}})g_4^2} \propto \frac{\varepsilon^2 \, e^2}{g_4^2},
\end{equation}
where we rescaled the $U(1)'$ charges and coupling such that $\text{max}(F_{\psi_{L/R}}) = 1$. Then, to create the desired value \eq{cp} of $a^d_{\pi^+ \pi^-}$ it follows from \eq{g4mz} that $g_4 \gtrsim \mathcal{O}(1)$ for the $Z^\prime$ mass range given by \eq{searchrange}. 
Putting everything together 
we obtain
$\kappa \lesssim 10^{-7}$. This provides an order of magnitude estimate for the branching ratios into $U(1)'$-uncharged fermions.

\section{High energy behavior \label{sec:hi}}

Let us investigate the consequences of \eqref{eq:g4mz} for the consistency of the $U(1)^\prime$  gauge group at higher energies. We begin asking about perturbativity. Using 
\begin{align}
  g_4\,F_\psi \lesssim  4 \pi\,,  
\end{align}
for all SM fields $\psi$ the $Z^\prime$ mass gets bounded from above
\begin{align}
    M_{Z^\prime}\lesssim 400\,\text{GeV}~.
\end{align}

Next, we investigate the occurrence of Landau poles (LP). 
Neglecting kinetic mixing effects,  
the scale $\mu_\text{LP}$ of the $U(1)^\prime$ pole can be estimated as
\begin{align}
\mu_\text{LP} =\mu \cdot  \text{exp}\left[ \frac{(4 \pi)^2}{g_4^2(\mu) B_4}\right]
\end{align}
with the one-loop coefficient in the beta-function
\begin{equation} \label{eq:B4}
    B_4=\sum_\psi \frac{4}{3} d_2(\psi) d_3(\psi) F_\psi^2\,,
\end{equation}
counting over all Weyl fermions $\psi$. 
The corresponding gauge coupling $g_4$ is fixed by data, e.g. \eqref{eq:g4mz}, at the scale of NP $\mu = M_{Z^\prime}$. One obtains schematically
\begin{equation}
    \frac{g_4(M_{Z'})}{M_{Z'}} = \frac{1}{\sqrt{B_4}\,m_0}\,,
\end{equation}
with $m_0$ being a characteristic mass scale. Thus the scale of the Landau pole is around 
\begin{align} \label{eq:pole}
\mu_\text{LP} = M_{Z^\prime}  \cdot  \text{exp}\left[ \frac{(4 \pi)^2 m_0^2}{ M_{Z^\prime}^2}\right]\,.
\end{align}
For large $Z^\prime$ masses $M_{Z^\prime} \gg 4 \pi \,m_0 $ this implies   $\mu_\text{LP} =M_{Z^\prime}$. 
Nearby  Landau poles can hence only be avoided  if  $M_{Z^\prime} \ll 4 \pi \,m_0$. 

\begin{figure}
    \centering
    \includegraphics[width=0.95\columnwidth]{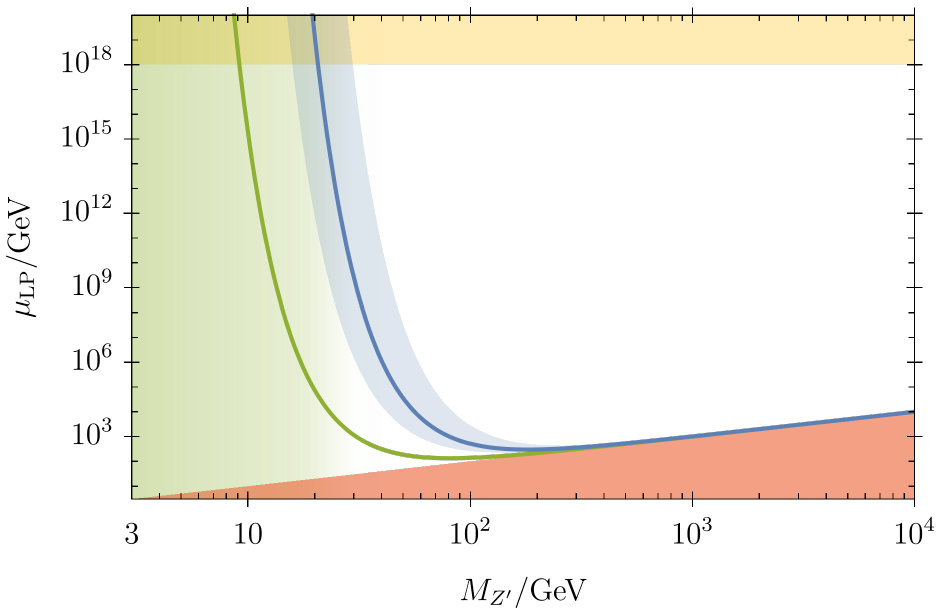}
    \caption{Scale of the Landau pole $\mu_\text{LP}$ via the estimate \eqref{eq:pole} for BM~III (blue), depending on the $M_{Z'}$ mass. 
    An uncertainty of $30\%$ is considered for $a^d_{\pi^- \pi^+}$ in \eqref{eq:g4mz}, consistent with \eqref{eq:cp} (blue shaded area).
    The shifted central value is also shown if one dark fermion $F_\chi = 4\,F$ is included (solid green line).
    The red shaded area is excluded, as $\mu_\text{LP} \leq M_{Z'}$.
    The preferred mass range \eqref{eq:searchrange} of $M_{Z'}$ is shaded in green. The yellow band indicates the $\mu_\text{LP}$ regime an order of magnitude around the Planck scale.   
     }
    \label{fig:LP}
\end{figure}

 In \fig{LP}, the location of the Landau pole with respect to $M_{Z^\prime}$  is illustrated   for BM~III, where $B_4 = \tfrac{32}3 F^2$ and \eq{BMIII-constraint-g4} imply $m_0 \approx 10$~GeV. 
 Shown are the central value for $\mu_{\rm LP}$ (blue line), a $30\%$ uncertainty band for $a^d_{\pi^- \pi^+}$ in \eqref{eq:g4mz}, consistent with \eqref{eq:cp} (blue shaded area), and the shifted central value due to a dark fermion (with $F_\chi = 4\,F$, solid green line)~\footnote{With $N_\chi$ additional  vector-like, $U(1)'$-charged dark fermions $\chi$, see Sec. \ref{sec:Zp-decay}, $B_4$ increases by $\delta B_4=\tfrac83 F_\chi^2 N_\chi$.}.
The red shaded area ($\mu_\text{LP} \leq M_{Z'}$) is excluded.
 We observe that the Landau pole may be beyond the planckian regime (yellow band) only for very light $Z'$ ($M_{Z'} \lesssim 20$~GeV, green shaded area) and subplanckian otherwise. Increasing $M_{Z'}$, the Landau pole  decreases towards a TeV-ish minimum (where $M_{Z'} \gtrsim 81$~GeV), implying an upper bound for $M_{Z'}$. For $M_{Z'}$ beyond the electroweak scale, the Landau pole stays exponentially close to $M_{Z'}$, indicating that the theory is strongly coupled. Finally, we note that the addition of extra $U(1)'$ charge carriers, $e.g.$~a dark fermion, shifts the Landau pole towards lower values, and more substantially so for low $M_{Z'}$.

This overall pattern is in sharp contrast to the situation observed in rare $b$-decays, where the anomalies indicate $\sqrt{B_4}\,m_0 \simeq 40$ TeV and a TeV-size $Z^\prime$ is vastly separated from the Landau pole. 
This permits the introduction of
additional  BSM-interactions to evade the Landau pole, and to secure
Higgs stability all the way up to the Planck scale (and possibly beyond) \cite{Bause:2021prv}.

Finally, we  emphasize that a sub-TeV-ish Landau pole is avoided as long as 
\begin{align} \label{eq:LP-target}
    M_{Z^\prime} \lesssim {\text{few}}  \times 10 \,  \text{GeV}~,
\end{align}
consistent with search limits \eqref{eq:searchrange} and $q \bar q$-data (\ref{eq:upsilon}), (\ref{eq:det}).

\section{Conclusions \label{sec:con}}

The recent data on charm CP-violation  \eqref{eq:KKLHCB}, taken at face value, together with the measurement of $\Delta A_{\text{CP}}$~\eqref{eq:ACPLHCB}, require to accept  a huge amount of $U$-spin breaking, or NP below the weak scale.
Explaining~\refeq{cp} indeed poses a challenge to model building, given the low NP scale and the severe constraints from rare decays, mixing, and searches for BSM bosons in dilepton and dijets channels.

We obtain viable explanations of data \eqref{eq:cp} from a light $Z'$-boson, $M_{Z^\prime} < \text{few}  \times 10  \,  \text{GeV}$, with novel, characteristic patterns in couplings to fermions: successful model benchmarks, see Tab.~\ref{tab:ZprimeModels}, accidentally couple only to right-handed fermions and are leptophobic (BM IV) or do not have $U(1)^\prime$-charges to electrons or muons (BM III). The latter is subject to kinetic mixing at one loop, which requires a non-excessive tuning of the mixing parameter at the level of ten percent. The former is a novel benchmark using a Diophantine construction to maintain anomaly freedom and the absence of one-loop kinetic mixing, since it couples mostly to quarks without introducing new matter fields, derived in App.~\ref{sec:newBM}.

Models do not couple to top quarks allowing for the top Yukawa to be written down directly. BM III  allows also the bottom Yukawa at tree level. We stress that we are not addressing the origin of flavor, which is beyond the scope of this work.
Models can also include a dark sector, which can significantly speed up the running such that low scale Landau poles arise which point to a UV-completion as low as a TeV,
see Fig.~\ref{fig:LP}.
However, in view of  the "the lighter, the safer"-rule a sufficiently light $Z^\prime$, with details depending on the dark sector, can avoid this.

Model frameworks BM III and BM IV are unique as they are minimal models  passing 
the very many  theoretical and experimental constraints. Together with viable variants obtained by swapping charges within one species, such as BM III-s, they are indeed the only, minimal options, inducing CP-violation in the  $\bar c_R u_R$ current and an enhanced $\bar d_R d_R$ current,  but it cannot be ruled out that further,  highly tuned scenarios with cross talk between species may be constructed.
Another  intriguing feature of the models  is  that  they can simultaneously explain the charm CP-data \eqref{eq:cp} and the 
$J/\psi \to  \pi^+ \pi^-$, $\psi^\prime \to  \pi^+ \pi^-$  branching ratios for a $Z^\prime$ around $\sim 3$ GeV (BM III) or  $\sim (5-7)$ GeV (BM IV), see Sec.~\ref{sec:quark}, 
providing a NP-explanation to  the  long-standing pion form factor puzzle.

Models can be searched for in low mass dijets along the lines of \cite{CMS:2019xai} or $\Upsilon$ and $J/\psi$ or $\psi^\prime$-decays and related dark photon searches.
The dominant $Z^\prime$ branching ratios are given in Tab.~\ref{tab:Brs}.
Signatures include enhanced production in   $\pi^+ \pi^-$, or $DD$, and $\tau \tau$ in BM III.
 If dark fermions are also present signatures as in Fig.~\ref{fig:missing} with hadrons and invisibles are promising smoking guns.
 We stress  that BM III-s, a variant of BM III  with the  charges between first and second generation up-type quark singlets flipped, $F_{u1}=-F$, $F_{u2}=G$ is equally  viable.
It has a different phenomenology than the other benchmark models, as it does couple essentially to first generation quarks, the $\tau$, and neutrinos, hence evades charmonium limits, and could be  as light as ${\cal{O}}( \text{GeV})$ or possibly below.

All viable models  further predict isospin violation and pattern in the CP-asymmetries in hadronic charm decays, see \eqref{eq:correlation}, in 
addition to $A_{\text CP} (D \to K_S K_S)$, which also requires U-spin breaking \cite{Grossman:2006jg}.
Models are tightly constrained by $D$-mixing and $Z^\prime$-searches into electrons and muons.
They can hence signal NP in the next round of data.

\acknowledgments
 
We are happy to thank Tommaso Pajero and Stefan Schacht for useful communication, and Alex Gilman for pointing out  charmonia bounds for taus and invisibles. This work is supported by the \textit{Studienstiftung des Deutschen Volkes} (TH), the \textit{Bundesministerium f\"ur Bildung und Forschung} (BMBF) under project number 05H21PECL2 (HG), and the Science and Technology Research Council (STFC) under the Consolidated Grant ST/T00102X/1 (DFL). This work was  performed in part at Aspen Center for Physics (GH, DFL), which is supported by National Science Foundation grant PHY-1607611, and was partially supported by a grant from the Simons Foundation (DFL).
GH is grateful to the MITP of the PRISMA$^+$ Cluster (Project ID 39083149), for its hospitality and partial support during the completion of this work.

\appendix 

\section{Amplitudes and observables in SM \label{app:SM}}

The  SM decay amplitudes of 
$D \to \pi^+ \pi^-$ and its conjugate decay  can be written in terms of reduced amplitudes $t,h$
with relative strong phase $\delta$ as 
\begin{align}
A=A(D^0 \to \pi^+ \pi^-)^{\text{SM}}&= \Sigma  \,  t  + V_{cb}^* V_{ub} h e^{i \delta} \, , \\
\bar A= A(\bar D^0 \to \pi^+ \pi^-)^{\text{SM}}&= \Sigma^*  t  + V_{cb} V_{ub}^* h e^{i \delta} \,  ,
\end{align}
where $\Sigma= (V_{cd}^* V_{ud}-V_{cs}^* V_{us})/2$ and  $V_{cs}^* V_{us} \simeq -V_{cd}^* V_{ud}$ due to CKM-unitarity up to a negligible  $V_{cb}^* V_{ub}$.
The amplitude $t$ is predominantly induced by tree-level $W$-exchange, whereas $h$ originates from
higher order contributions including penguin-loops or final-state interactions.
Using standard formulas one obtains for the direct 
CP-asymmetry of $D \to \pi^+ \pi^-$ decays in the SM 
\begin{align} \label{eq:acpSM}
 a_{\pi^-\pi^+}^{d \, \text{SM}} &= {\frac{ |A|^2- |\bar A|^2}{ |A|^2+|\bar A|^2}} \simeq  2 \cdot \text{Im}( \frac{V_{cb}^* V_{ub}}{V_{cd}^* V_{ud}}) \frac{h}{t} \sin \delta \, , 
\end{align}
where we neglected contributions of higher order in  CKM-ratios $V_{cb}^* V_{ub}/(V_{cd}^* V_{ud}) \sim 10^{-3}$. This approximation implies also that the CP-averaged  branching ratio
is dominated by the term with the leading CKM-factor $\Sigma$, ${\cal{B}} (D \to  \pi^+  \pi^-) \propto  |A|^2+|\bar A|^2 \propto |\Sigma|^2  t^2$.

Expressions for  $D \to K^+  K^-$ decays are analogous:
\begin{align}
A(D^0 \to K^+  K^-)^{\text{SM}}&= -\Sigma \,  t_s  + V_{cb}^* V_{ub} h_s e^{i \delta_s} \, , \\
 A(\bar D^0 \to K^+  K^-)^{\text{SM}}&= -\Sigma^*  t_s  + V_{cb} V_{ub}^* h_s e^{i \delta_s} \, . 
\end{align}
 In the $U$-spin limit  the phase and reduced amplitudes  are universal, such as  $t=t_s=\bar t$ and so on.
 The observed $U$-spin breaking  in the branching ratios   ${\cal{B}} (D \to  K^+  K^-)/{\cal{B}} (D \to  \pi^+  \pi^-)\simeq 2.8$
can be explained with a flavor-dependent correction $\delta t$, as  $t_s=\bar t  +\delta t$, $t=\bar t - \delta t$
 at the nominal level of flavor breaking, $\delta t/\bar t \simeq 30 \%$ \cite{Brod:2012ud}.
 This  $U$-spin breaking in the branching ratios and correspondingly  between the leading contributors to the amplitudes $t,t_s$ implies a shift in the CP-asymmetries e.g.~(\ref{eq:acpSM}) 
 and leads to 
  the modified  $U$-spin relation (\ref{eq:modified}).

\section{RGE and hadronization}\label{app:A}

NP effects in charm decays are described within the effective weak Hamiltonian $\mathcal{H}_{\text{eff}}^{|\Delta \text{c}|=1}\:\supset\:\frac{G_F}{\sqrt{2}}\,\sum_i\,\widetilde{C}_i^{(\prime)}\,\widetilde{Q}_i^{(\prime)}$. 
Operators relevant to a $Z^\prime$-boson coupling to right-handed quarks read
\begin{align}
\widetilde{Q}_9^\prime &= (\bar u c)_{V+A} \sum_{q} F_{u_i,d_i}\, (\bar q q)_{V+A}~,\\
    \widetilde{Q}_{10}^{\prime} &= (\bar u_\alpha c_\beta)_{V+A} \sum_{q}F_{u_i,d_i}\, (\bar q_\beta q_\alpha)_{V+A}~,\label{eq:OpZ10}
\end{align}
where $(V\pm A)$ refers to the Dirac structures $\gamma_\mu (1 \pm \gamma_5)$, $q=u,c,d,s,b$, 
and $\alpha,\beta$ are color indices.

In the following, 
we address the evolution of the Wilson coefficients at the $Z^\prime$ mass scale,
\begin{align}\label{eq:WCs}
    \widetilde{C}_{9}^\prime(M_{Z^\prime})=\frac{\sqrt{2}}{G_F} \Delta \tilde F_R \frac{g^2_4}{4\,M_{Z^\prime}^2} e^{i \Phi_R}\, ,  \widetilde{C}_{10}^\prime(M_{Z^\prime})=0 \, , 
\end{align}
down to the charm mass scale at  leading order in $\alpha_s$, see~\cite{Bause:2020obd}  for  details. 
The CP-violating phase is $\Phi_R \sim \pi/2$. 
Using the anomalous dimension given by Eq.~(B1) in Ref.~\cite{Bause:2020obd}, and integrating out degrees of freedom at the ${(Z^\prime, t, b)}$--scales, one obtains
\begin{align}\label{eq:evo}
    \vec{C}(\mu)\,=\,U(\mu,M_{Z^\prime})\,\vec{C}(M_{Z^\prime})~,
\end{align}
where $U(m_1,m_2)$ is the evolution matrix from scale $m_2$ to scale $m_1$. Using Eq.~\eqref{eq:WCs} and \eqref{eq:evo}, we obtain
\begin{align}
    \widetilde{C}_{9}^{\prime}(m_c)&=\frac{1}{2}\left(R^{\frac{1}{2}}+R^{-1}\right)\,\widetilde{C}_{9}^{\prime}(M_{Z^\prime})~,\\
    \widetilde{C}_{10}^{\prime}(m_c)&=\frac{1}{2}\left(R^{\frac{1}{2}}-R^{-1}\right)\,\widetilde{C}_{9}^{\prime}(M_{Z^\prime})~,
\end{align}
where
\begin{align}
    R=\bigg(\frac{\alpha_s^{(4)}(m_b)}{\alpha_s^{(4)}(m_c)}\bigg)^{\frac{12}{25}}\bigg(\frac{\alpha_s^{(5)}(m_t)}{
  \alpha_s^{(5)}(m_b)}\bigg)^{\frac{12}{23}} \bigg(\frac{\alpha_s^{(6)}(M_{Z^\prime})}{\alpha_s^{(6)}(m_t)}\bigg)^{\frac{4}{7}}~.
\end{align}
for $M_{Z^\prime}> m_t$, while
\begin{align}
    R=\bigg(\frac{\alpha_s^{(4)}(m_b)}{\alpha_s^{(4)}(m_c)}\bigg)^{\frac{12}{25}}\bigg(\frac{\alpha_s^{(5)}(M_{Z^\prime})}{
  \alpha_s^{(5)}(m_b)}\bigg)^{\frac{12}{23}}~,
\end{align}
for $m_b\leq M_{Z^\prime}\leq m_t$.

\begin{figure}[t]
    \centering
    \includegraphics[width=0.45\textwidth]{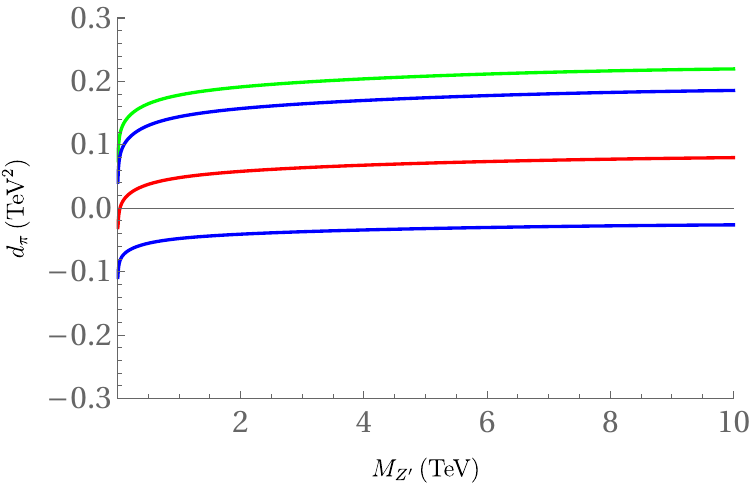}
    \includegraphics[width=0.45\textwidth]{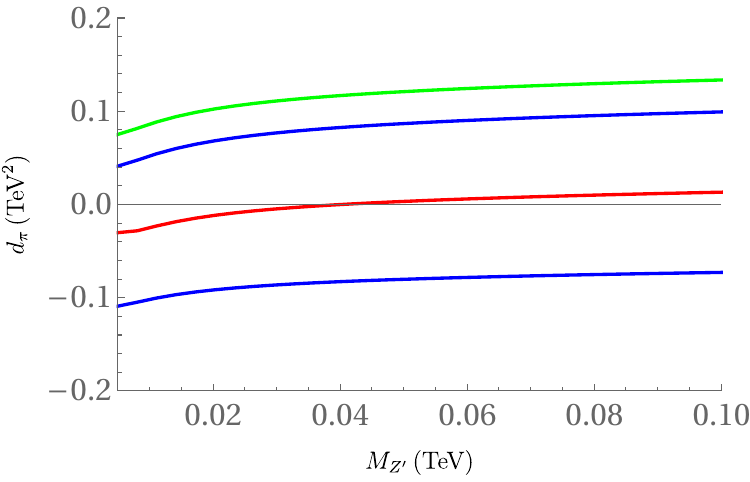}
    \caption{$d_\pi$ as a function of $M_{Z^\prime}$ as in \eqref{eq:dpi}, \eqref{eq:r2}. The  red line is obtained for $B_{9,10}^{\pi^+ \pi^-}=1$, that is the naïve factorization approach. 
    Here, $d_\pi$ crosses zero around 40 GeV. Blue lines account for a deviation of 30\% from the naïve factorization limit, that is $B_{9,10}^{\pi^+ \pi^-}=1\pm0.3$. The green line represents the large-$N_C$ limit.}
    \label{fig:dpiMzp}
\end{figure}

For the  computation of the $D^0\to K^+K^-$ and $D^0\to\pi^+\pi^-$ hadronic matrix elements we employ factorization of currents, 
\begin{align}
&\bra{P^+\,P^-} Q_i\ket{D^0}\\
&= \bra{P^+ }(\bar{q}_1\,\Gamma_1\, q_2)\ket{0}\,\bra{P^-}(\bar{q}_3\, \Gamma_2\, q_4)\ket{D^0}\,B^{P^+ P^-}_i~, \nonumber
\end{align}
with $P=\pi,K$, $Q_i=(\bar{q}_1\,\Gamma_1\,q_2)\,(\bar{q}_3\, \Gamma_2\, q_4)$ is a $4$--quark operator and $\Gamma_{1,2}$ represent Dirac and color structures while $q_j$ denote quarks. The factor $B^{P^+ P^-}_i$ parametrizes the deviation of the true hadronic matrix element from its naïve approximation, 
$B^{P^+ P^-}_i \vert_{\text{naïve}}=1$. Including these effects the CP-asymmetries become 
\begin{align}
\begin{split}
    a_{K^-K^+}^d&\propto F_{d_2} \left( \frac{\widetilde{C}_{9}^{\prime}(m_c)}{N_C} B_9^{K^-K^+}+\widetilde{C}_{10}^{\prime}(m_c)\,B_{10}^{K^-K^+}\right),\\
    a_{\pi^-\pi^+}^d&\propto F_{d_1}\left( \frac{\widetilde{C}_{9}^{\prime}(m_c)}{N_C}B_9^{\pi^-\pi^+}+\widetilde{C}_{10}^{\prime}(m_c)B_{10}^{\pi^-\pi^+}\right),
\end{split}
\end{align}
where we have already used $F_{Q_{1,2}}=0$. Furthermore
\begin{align}
\begin{split} \label{eq:dpi}
        d_K\,&=\,\frac{1}{a_K}\,r_2^K(m_c,M_{Z^\prime})~,\,\,d_\pi\,=\,-\,\frac{1}{a_\pi}\,r_2^\pi(m_c,M_{Z^\prime})~,
\end{split}
\end{align}
with 
\begin{align}\label{eq:r2}
     r_2^P(m_c,M_{Z^\prime})&= \frac{B_+^P\,R^{1/2}-B_-^P\,R^{-1}}{\sqrt{2}\,G_F\,\lambda_s}~,
\end{align}
and
\begin{align}
    B_\pm^P\,=\,\frac{1}{2}\,\left(B_{10}^{P^+ P^-}\pm \frac{1}{N_C}\,B_9^{P^+ P^-}\right)~.
\end{align}
In  naïve factorization, $B_+^P=2/3$ and $B_-^P=1/3$, recovering Eq.~(D2) from Ref.~\cite{Bause:2020obd}. In the large-$N_C$ limit $B_\pm^P=(1/2) B_{10}^{P^+ P^-}$. We show $d_\pi$ against $M_{Z^\prime}$   in Fig.~\ref{fig:dpiMzp}. In  naive factorization (red lines) $d_\pi$  exhibits a strong cancellation around $\sim 40$ GeV stemming from the numerator of Eq.~\eqref{eq:r2}. Comparing with the large-$N_C$ limit (green lines), we observe that this cancellation is only effective in the naïve factorization limit. Taking into account that naïve factorization suffers from sizable uncertainties, blue lines illustrate $30\%$, in our analysis for light $Z^\prime$ masses we  use  $d_\pi\simeq 0.1\,$TeV$^2$.  

\section{Anomaly cancellation \label{sec:newBM}}

The gauge anomaly cancellation conditions (ACCs) read~\cite{Allanach:2018vjg,Bause:2019vpr}
{\small
\begin{align}
    &2\langle\mathcal{F}_Q\rangle-\langle\mathcal{F}_u\rangle-\langle\mathcal{F}_d\rangle=0\,,\label{eq:ACC1}\\
    &3\langle\mathcal{F}_Q\rangle+\langle\mathcal{F}_L\rangle=0\,,\label{eq:ACC2}\\
    &\langle\mathcal{F}_Q\rangle+3\langle\mathcal{F}_L\rangle-8\langle\mathcal{F}_u\rangle-2\langle\mathcal{F}_d\rangle-6\langle\mathcal{F}_e\rangle=0\,,\label{eq:ACC3}\\
    &6\langle\mathcal{F}_Q\rangle+2\langle\mathcal{F}_L\rangle-3\langle\mathcal{F}_u\rangle-3\langle\mathcal{F}_d\rangle-\langle\mathcal{F}_e\rangle-\langle\mathcal{F}_\nu\rangle=0\,,\label{eq:ACC4}\\
    &\langle\mathcal{F}_Q^2\rangle-\langle\mathcal{F}_L^2\rangle-2\langle\mathcal{F}_u^2\rangle+\langle\mathcal{F}_d^2\rangle+\langle\mathcal{F}_e^2\rangle=0\,,\label{eq:ACC5}\\
    &6\langle\mathcal{F}_Q^3\rangle+2\langle\mathcal{F}_L^3\rangle-3\langle\mathcal{F}_u^3\rangle-3\langle\mathcal{F}_d^3\rangle-\langle\mathcal{F}_e^3\rangle-\langle\mathcal{F}_\nu^3\rangle=0\,.\label{eq:ACC6}
\end{align}}In addition, avoiding $Z^\prime-Z$ kinetic mixing at one loop requires~\cite{Holdom:1985ag,Bause:2020obd}
\begin{align}
    \langle\mathcal{F}_Q\rangle-\langle\mathcal{F}_L\rangle+2\langle\mathcal{F}_u\rangle-\langle\mathcal{F}_d\rangle-\langle\mathcal{F}_e\rangle=0\,.\label{eq:ACC7}
\end{align}
Here, we use the notation $\langle X\rangle=\text{Tr}(X)$ and $\mathcal{F}_A=\text{diag}(F_{A_1},F_{A_2},F_{A_3})$ with $A=Q,u,d,L,e,\nu$. First, we focus on those equations that are linear with trace charge matrices, that are Eqs.~\eqref{eq:ACC1},\eqref{eq:ACC2},\eqref{eq:ACC3},\eqref{eq:ACC4}, and \eqref{eq:ACC7}. Solving them, we find
{\small
\begin{align}\label{eq:tracerelation}
    &\langle\mathcal{F}_Q\rangle=
    -\langle\mathcal{F}_u\rangle=
    \frac{1}{3}\langle\mathcal{F}_d\rangle\nonumber=
    -\frac{1}{3}\langle\mathcal{F}_L\rangle=
    -\langle\mathcal{F}_e\rangle=
    -\frac{1}{5}\langle\mathcal{F}_\nu\rangle\,.
\end{align}
}
Setting  $F_{Q_{1,2,3}}=0$ we avoid kaon constraints and for the BMs III and IV couplings to the top, and arrive at the simple condition
\begin{align}
    \langle\mathcal{F}_Q\rangle=0\,,
\end{align}
and therefore from Eq.~\eqref{eq:tracerelation} to
\begin{align}\label{eq:tracezero}
    \langle\mathcal{F}_u\rangle=
    \langle\mathcal{F}_d\rangle=
    \langle\mathcal{F}_L\rangle=
    \langle\mathcal{F}_e\rangle=
    \langle\mathcal{F}_\nu\rangle=0\,.
\end{align}
Let us work on the  remaining ACCs, Eqs.~\eqref{eq:ACC5} and \eqref{eq:ACC6}. The non-linear behavior of these equations makes it challenging to solve them. We have already some information about the trace of these matrices, they are zero~\eqref{eq:tracezero}. To solve the problem mathematical relations between $\langle\mathcal{F}_A\rangle$, $\langle\mathcal{F}_A^2\rangle$ and $\langle\mathcal{F}_A^3\rangle$ would be helpful. Noting that for $3\times 3$ matrices holds
\begin{align}\label{eq:cayley-hamilton}
    \langle\mathcal{F}_A^3\rangle-\frac{3}{2}\langle\mathcal{F}_A^2\rangle\langle\mathcal{F}_A\rangle+\frac{1}{2}\langle\mathcal{F}_A\rangle^3=
    3  \, \text{det}( \mathcal{F}_A )~,
\end{align}
and $\text{det} (\mathcal{F}_A)=  F_{A_1} F_{A_2} F_{A_3}$ it follows  from $\langle\mathcal{F}_A\rangle=0$ that $\langle\mathcal{F}_A^3\rangle$ vanishes if one charge vanishes, for instance,  $F_{A_3}=0$; then Eq.~\eqref{eq:ACC6} is fulfilled. In general, the charge matrices $\mathcal{F}_A$ can then be written as
\begin{align}\label{eq:appendix-model-charges}
    \mathcal{F}_A=F_A\begin{pmatrix}
+1 & \phantom{-}0 & \phantom{+}0\\
\phantom{+}0 & -1 & \phantom{+}0\\
\phantom{+}0 & \phantom{-}0 & \phantom{+}0
\end{pmatrix},\,\,\,A=u,d,L,e,\nu\,,
\end{align}
where $F_A$ are integers. Note that the order of $(+1, -1, 0)$ can be changed for each species $A$ independently. It remains to solve  Eq.~\eqref{eq:ACC5}, which now simply reduces to
\begin{align}\label{eq:dio1}
    F_d^2+F_e^2=F^2_L+2\,F_u^2\,.
\end{align}
We can start exploring this equation by setting $F_e=F_L$. This solution is motivated from the phenomenological point of view because we need small values of lepton charges as well as being disconnected from quark charges. However in this case the system has only the trivial solution $F_u=F_d=0$ because $F_d^2=2\,F_u^2$ can not be fulfilled for integers. Let us simplify \eqref{eq:dio1} by setting $F_L=0$ which is motivated by the strong constraints on lepton couplings. We obtain
\begin{align}\label{eq:diophan}
    F_d^2+F_e^2=2F_u^2\,.
\end{align}
Two aspects are worth noting from this equation:
\begin{enumerate}[label=(\alph*)]
    \item For any Pythagorean triple $(m,n,p)$, {\it i.e.} integers which solve Pythagoras equation $m^2 + n^2 = p^2$, one can find integer triples $(F_d,F_e,F_u)$ which solve Eq.~\eqref{eq:diophan} by substituting $F_d=m+n$, $F_e=m-n$, and $F_u=p$. 
    \item Any solution $(F_d,F_e,F_u)$ to Eq.~\eqref{eq:diophan} can be parametrised by integers $(p,q,r)$
\begin{align}
    F_d &= p^2 + (-4\, r + 2\, q) p + 2\, r^2 - q^2,\label{eq:dio1a}\\
    F_e &= p^2 - 2 \,r^2 - q^2 - 2 \,p \,q + 4 \,q \,r,\label{eq:dio2a}\\
    F_u &= p^2 + 2 \,r^2 + q^2 - 2 \,p \,r - 2 \,q \,r.\label{eq:dio3a}
\end{align}
The first non-trivial solutions $(F_d,F_e,F_u)$ are found to be 
\begin{align}
(F_d,F_e,F_u)&=(1, 1, 1),\ (7, 1, 5),\ (17, 7, 13),\cdots
\end{align}
and so on, and integer multiples thereof. Note that each term can also have either sign, and that permutations between $F_d$ and $F_e$ are permitted.
\end{enumerate}
Although Eq.~\eqref{eq:diophan} gets fully solved by Eqs.~\eqref{eq:dio1a}-\eqref{eq:dio3a}, in the following, we illustrate a more practical approach that allows us to run directly into those solutions that accommodate the constraint given by Eq.~\eqref{eq:leptonlimit}. Let us rewrite \eqref{eq:diophan} as
\begin{align}\label{eq:sandwich}
    \langle F|\mathcal{J}| F\rangle=F_e^2\,,
\end{align}
with $|F\rangle=(F_u,F_d)$ and 
\begin{align}
  \mathcal{J}=\begin{pmatrix}
    2& 0\\
    0& -1
    \end{pmatrix}~.  
\end{align}
For illustration, we start with the trivial solution $(F_d,F_e,F_u)= (1,1,1)$. Using Eq.~\eqref{eq:sandwich}, we obtain
\begin{align}\label{eq:sandwich1}
    \langle F_0|\mathcal{J}| F_0\rangle=1\,,
\end{align}
with 
\begin{align}
    |F_0\rangle&=(1,1)\,.
\end{align}
Now, the question we should ask is: are there more solutions such as $F_e\ll F_u,F_d$? One possibility is if a transformation 
\begin{align}
    \mathcal{J}\longrightarrow \mathcal{J}^\prime=U^T\,\mathcal{J}\,U=\mathcal{J}~,
\end{align}
by a $2\times 2$ matrix with integer entries $U$, leads invariant Eq.~\eqref{eq:sandwich1} so that we can generate recursively solutions
\begin{align}\label{eq:travel}
    |F_i\rangle=(U)^i\,|F_0\rangle\,,
\end{align}
which could get enlarged while keeping $F_e$ fixed to $1$ for this particular case. This matrix needs to satisfy 
\begin{align}
    \begin{pmatrix}
    U_{11}& U_{12}\\
    U_{21}& U_{22}
    \end{pmatrix}^T\begin{pmatrix}
    2& 0\\
    0& -1
    \end{pmatrix}\begin{pmatrix}
    U_{11}& U_{12}\\
    U_{21}& U_{22}
    \end{pmatrix}=\begin{pmatrix}
    2& 0\\
    0& -1
    \end{pmatrix}~.
\end{align}
The solution of this system in terms of $U_{22}$ is
\begin{align}
  U_{11}\,&=\,\eta_{11}\,U_{22}~,\\
U_{12}\,&=\,\eta_{12}\,\sqrt{\frac{-1+U_{22}^2}{2}}~,\\
U_{21}\,&=\,\eta_{21}\,\sqrt{2\,(-1+U_{22}^2)}~,
\end{align}
with four solutions $(\eta_{11},\eta_{12},\eta_{22})=(-1,-1,+1)$, $(-1,+1,-1)$, $(+1,-1,-1)$, $(+1,+1,+1)$. The smallest integer solution reads 
\begin{align}
    U=\begin{pmatrix}
    \eta_{11}\,3& \eta_{12}\,2\\
    \eta_{21}\,4& \eta_{22}\,3
    \end{pmatrix}\,.
\end{align}
We then find the solutions \eqref{eq:travel} for any integer $i>0$. Using $\eta_{12}=\eta_{22}=\eta_{11}=\eta_{21}=+1$ leads to
\begin{align} \nonumber
   |F_1\rangle&
   =\,(5,7)\,,\\ \nonumber
   |F_2\rangle&
   =\,(29,41)\,,\\ 
   |F_3\rangle&
   =\,(169,239)\,,\\ \nonumber
   |F_4\rangle&
   =\,(985,1393)\,,\\ \nonumber
   |F_5\rangle&
   =\,(5741,8119)\,,\\ \nonumber
   &\ \vdots
\end{align}
We learn that in order to avoid electron and muon constraints of Eq.~\eqref{eq:leptonlimit}, we need  solutions $|F_i\rangle$ with $i\geq 4$. Note that we could have also used 
a different  
setup for $\eta_{11},...$ to get different solutions. In addition, we can also find more solutions by choosing another initial integer triple $(F_d,F_u,F_e)$. 

In the phenomenological analysis, we considered the solution $i=4$ for BM IV, 
\begin{align}\label{eq:BMVpart}
    (F_d,F_e,F_u)=(1393,1,985)
\end{align}
with zero neutrino charges $F_\nu=0$. \\

\section{Kinetic mixing \label{sec:mix}}

In this appendix we discuss the naturalness of the gauge-kinetic mixing, occurring due to the parameter $\eta$ as in \eqref{eq:gauge-kinetic-mixing} between the $Z'$ and the photon, or equivalently the $Z'$ and hypercharge gauge boson before electroweak symmetry breaking. 
Ideally, one would like the RG evolution of this parameter to be technically natural, which means
\begin{align} \label{eq:nature}
    \frac{\mathrm{d} \eta}{\mathrm{d} \ln \mu } \propto \eta\,.
\end{align}
This would imply that kinetic mixing can naturally be switched off at all scales, or made to remain arbitrarily small.

For theories with the charge configuration \eqref{eq:appendix-model-charges}, as well as the Higgs not carrying a $U(1)'$ charge, 
kinetic mixing is natural at one-loop order. 
If only gauge contributions are taken into account, the naturalness remains intact even at higher loops, which is the result of the symmetry implied by \eqref{eq:appendix-model-charges}.
We have verified that this is the case until four loops, using the results of~\cite{Bednyakov:2021qxa,Davies:2021mnc}.

Starting at two-loop order, Yukawa interactions violate the naturalness \eqref{eq:nature}, unless they retain certain flavor textures. 
If the top quark does not carry $U(1)'$ charge as in BM~IV, the naturalness-violating terms do not feature the top Yukawa coupling at two loops, and are either suppressed by the other, much smaller Yukawas and/or loop factors.
Thus, these terms are negligible for the running of $\eta$, which becomes effectively natural.

However, the naturalness is broken at one loop below the scale where the first field carrying $U(1)'$ charge is integrated out, i.e. the bottom quark in BM~IV.

\end{document}